\documentclass[aps,prb,twocolumn,superscriptaddress,longbibliography]{revtex4-2}
\usepackage{graphicx}
\usepackage{bm}
\usepackage{natbib}
\usepackage{color}
\usepackage{epstopdf}
\usepackage{amsmath}
\usepackage{url}
\usepackage{amssymb}
\usepackage{braket}
\definecolor{darkGreen}{RGB}{0,110,0}
\definecolor{darkBlue}{RGB}{5,4,170}
\usepackage[colorlinks=true, urlcolor=darkBlue,citecolor=darkBlue,linkcolor=darkBlue]{hyperref}
\usepackage{pifont}
\usepackage{gensymb}

\begin{document}

\title{Resilience of topological superconductivity under particle current}

\author{Alfonso Maiellaro}
\affiliation{Dipartimento di Fisica "E.R. Caianiello", Universit\`{a} degli Studi di Salerno, Via Giovanni Paolo II, 132, I-84084 Fisciano (SA), Italy}
\author{Francesco Romeo}
\affiliation{Dipartimento di Fisica "E.R. Caianiello", Universit\`{a} degli Studi di Salerno, Via Giovanni Paolo II, 132, I-84084 Fisciano (SA), Italy}
\affiliation{INFN, Sezione di Napoli, Gruppo collegato di Salerno, Italy}
\author{Fabrizio Illuminati}
\affiliation{INFN, Sezione di Napoli, Gruppo collegato di Salerno, Italy}
\affiliation{Dipartimento di Ingegneria Industriale, Universit\`{a} degli Studi di Salerno, Via Giovanni Paolo II, 132, I-84084 Fisciano (SA), Italy}
\author{Roberta Citro}
\affiliation{Dipartimento di Fisica "E.R. Caianiello", Universit\`{a} degli Studi di Salerno, Via Giovanni Paolo II, 132, I-84084 Fisciano (SA), Italy}
\affiliation{INFN, Sezione di Napoli, Gruppo collegato di Salerno, Italy}

\date{December 9, 2022}

\begin{abstract}
We investigate the robustness of topological superconductors under the perturbing influence of a finite charge current. To this aim, we introduce a
modified Kitaev Hamiltonian parametrically dependent on the quasiparticle momentum induced by the current. Using different quantifiers of the topological phase, such as the Majorana polarization and the edge state quantum conditional mutual information, we prove the existence of a finite critical value of the quasiparticle momentum below which edge modes and topological superconductivity survive. We also discuss how a finite current breaks time reversal symmetry and changes the topological class in the Altland-Zirnbauer classification scheme compared to the case of isolated systems. Our findings provide a nontrivial example of the interplay between topology and the nonequilibrium physics of open quantum systems, a relation of crucial importance in the quest to a viable topological quantum electronics.
\end{abstract}

\maketitle

\section{Introduction}
\label{introduction}
In the last two decades topological superconductivity has attracted a steadily growing interest, not least due to its potential role in conceiving innovative devices of quantum electronics. The simplest model of topological superconductivity was proposed by Kitaev in 2001 \cite{Kitaev_2001}. It consists of a one-dimensional spinless $p$-wave superconductor in which Majorana bound states (MBSs) are pinned to zero energy and localize at the edges. Indeed, an effective $p$-wave pairing can be realized by proximizing  semiconducting nanowires to s-wave superconductors \cite{PhysRevLett.100.096407}. Having thus some well identified condensed-matter physical counterparts \cite{PhysRevLett.105.177002,PhysRevLett.105.077001}, the Kitaev wire has become an established paradigm in studying the robustness of superconducting topological phases, as it allows to gain insight, with limited computational efforts, into the response of real devices to system modifications, material imperfections and environmental perturbations.  Accordingly, robustness of MBSs has been tested in the presence of imperfections \cite{PhysRevB.84.144526,PhysRevB.88.064506,PhysRevB.94.115166,Neven_2013}, multi-modes geometries \cite{PhysRevB.89.174514,Maiellaro2018TopologicalPD,PhysRevLett.105.227003,MaiellaroProc1,PhysRevB.96.035306,condmat6020015,condmat7010026}, and long-range hopping and/or pairing terms \cite{PhysRevB.97.041109,PhysRevB.88.165111,MaielTopTie,PhysRevLett.113.156402,PhysRevB.95.195160}. These studies have proved the resilience of topologically ordered phases against various realistic sources of perturbations, suggesting that superconducting topological order can be considered as a valuable resource in future and emerging quantum electronic technologies.

Most of the experimental efforts to detect emergent MBSs rely on metal/superconductor junctions \cite{PhysRevLett.98.237002,PhysRevLett.103.237001,PhysRevB.63.144531,nano9060894,PhysRevB.86.224511,PhysRevLett.119.136803,PhysRevLett.110.126406,PhysRevLett.109.267002} and Josephson junctions based on helical materials \cite{Kwon,PhysRevB.84.180502}. Indeed, once a current flux is injected into the systems, signatures of MBSs can be revealed by tunneling spectroscopy, via the zero-bias quantized peak, or by interferometric devices able to identify the $4\pi$-periodic Josephson effect. On the other hand, and quite crucially, the currents injected via source/drain terminals lead to undesired nonequilibrium effects on the topological phases, introducing a novel source of environmental perturbation. For this reason, despite the above-mentioned rich literature on isolated systems, it is particularly relevant to gain some understanding of  the interplay between nonequilibrium physics and topology for open systems in realistic conditions.

Recent works have approached the study of topological systems coupled to the evironment by imposing generalized boundary conditions \cite{PhysRevB.104.134516,PhysRevB.101.094502,PhysRevB.106.155407}. These methods, which share some similarity with previous investigations based on a self-energy approach \cite{Aguadoself,Maiellaro_2020}, incorporate information on the environment by emulating particle-hole symmetry breaking mechanisms originating from quasiparticle poisoning or boson-assisted tunnelling phenomena \cite{PHSb}. Such approaches are limited to situations where the net current injected into the system is negligible, so that they cannot be applied to important situations where current-induced nonequilibrium effects cannot be neglected. We are thus in need of effective models capable of incorporating genuine nonequilibrium features of open topological systems. While treating the full nonequilibrium dynamics of the system-environment interaction remains a formidable task, we expect that important information can be recovered by studying simple models that incorporate the effects of charged current flows.

In analogy with the Peierls substitution commonly used in modeling the influence of a vector potential in a tight-binding framework, the effect of a particle current on a superconducting Kitaev wire can be taken into account by introducing a phase factor $e^{iq}$ in the hopping integrals, where the wave vector $q$ quantifies the quasiparticle momentum induced by the current. In the presence of $p$-wave superconducting correlations, the complex phase induces a finite momentum $2q$ of the Cooper pair, parallel to the direction of the current. Accordingly, in the following we introduce such modified version of the Kitaev model and we investigate systematically the resilience of the edge modes by looking at different witnesses of the topological superconducting order, including the long-distance, edge-to-edge quantum conditional mutual information (QCMI) that measures the nonlocal correlations of the Majorana excitations \cite{PhysRevResearch.4.033088,MaieIllum2}, and the Majorana polarization of the zero-energy modes \cite{BENA2017349,PhysRevLett.110.087001,PhysRevB.85.235307,MaiellaroGeoFrust}.

\begin{figure*}
	\includegraphics[scale=0.11]{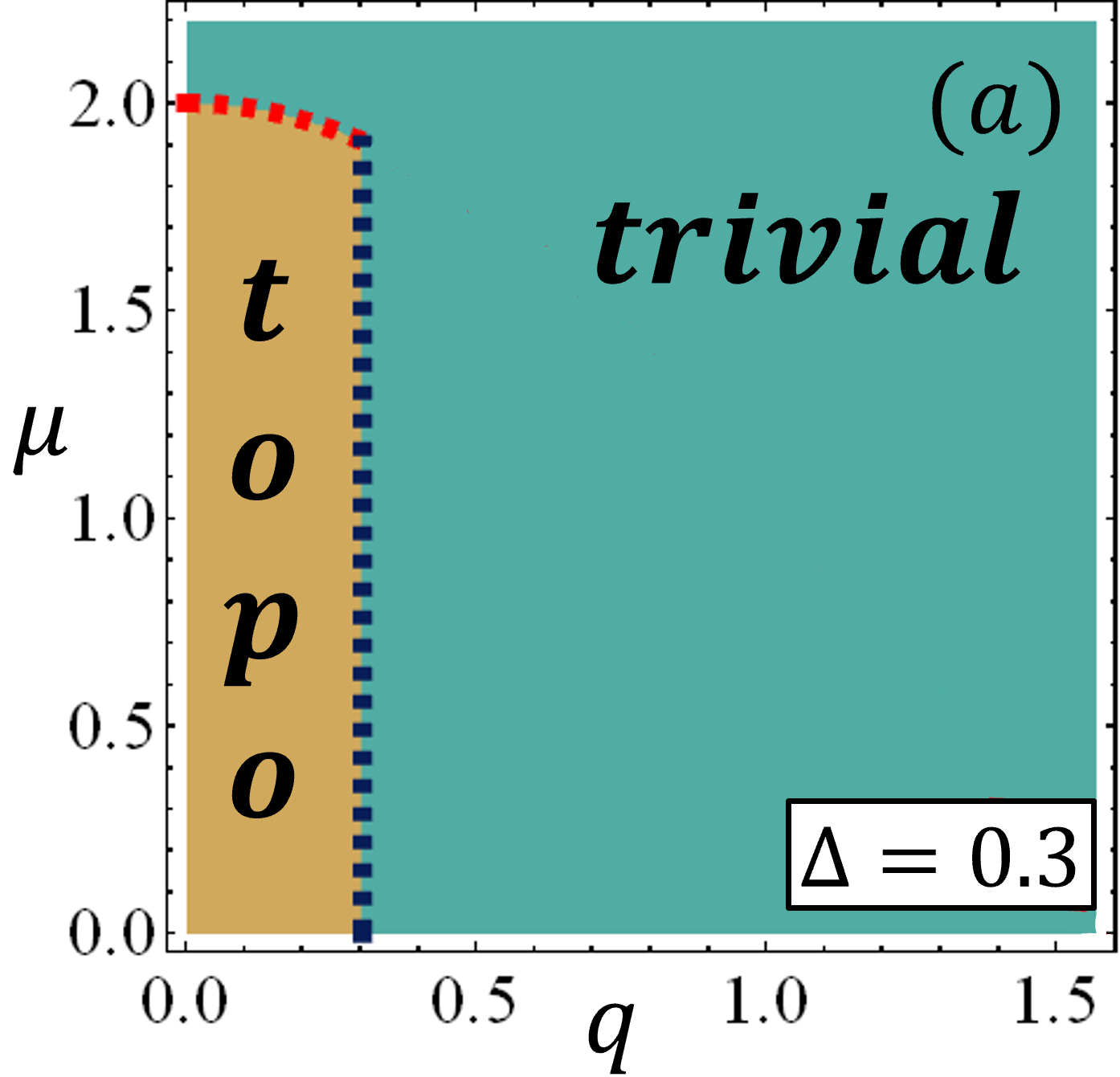}
	\includegraphics[scale=0.11]{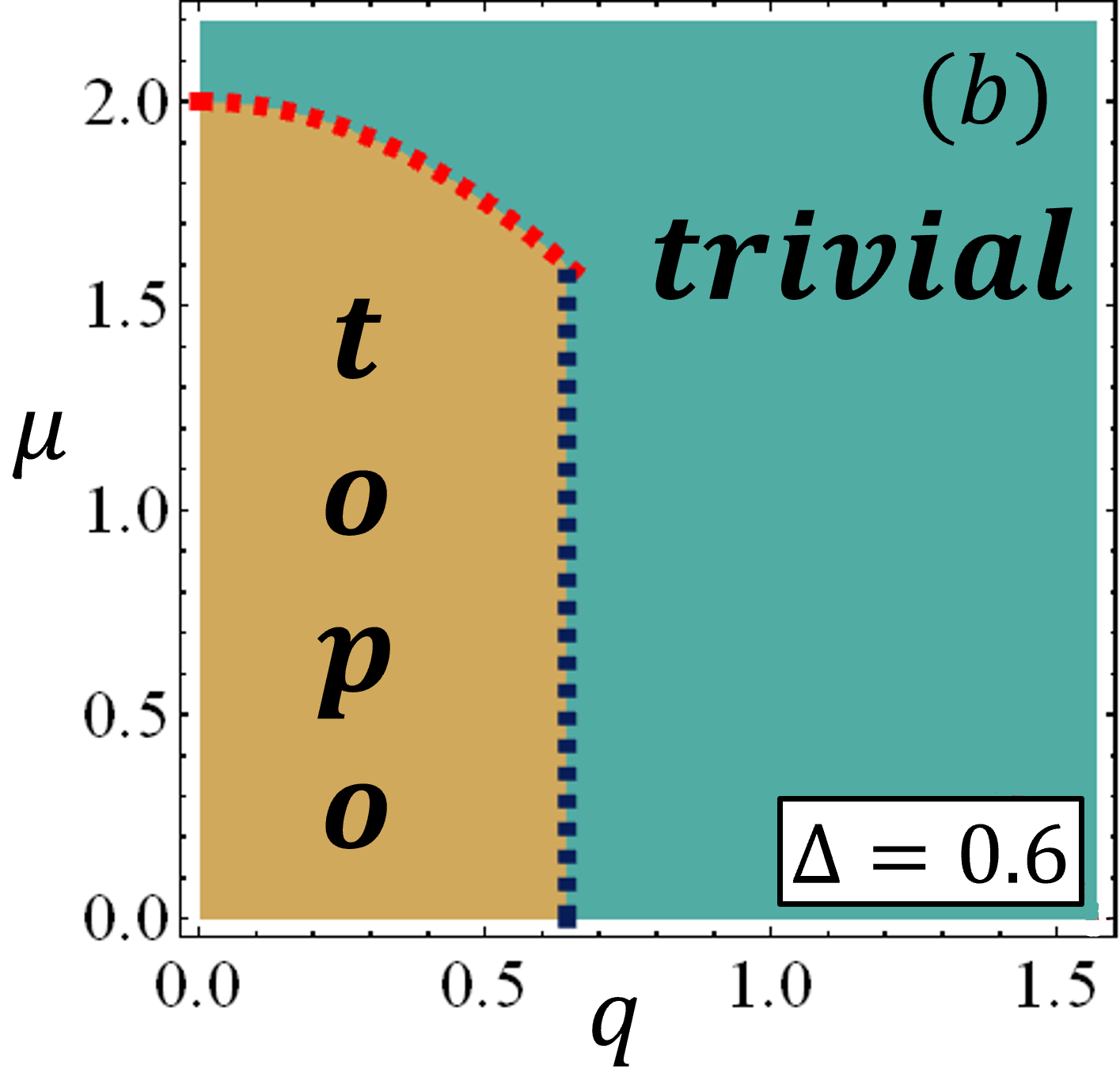}
	\includegraphics[scale=0.11]{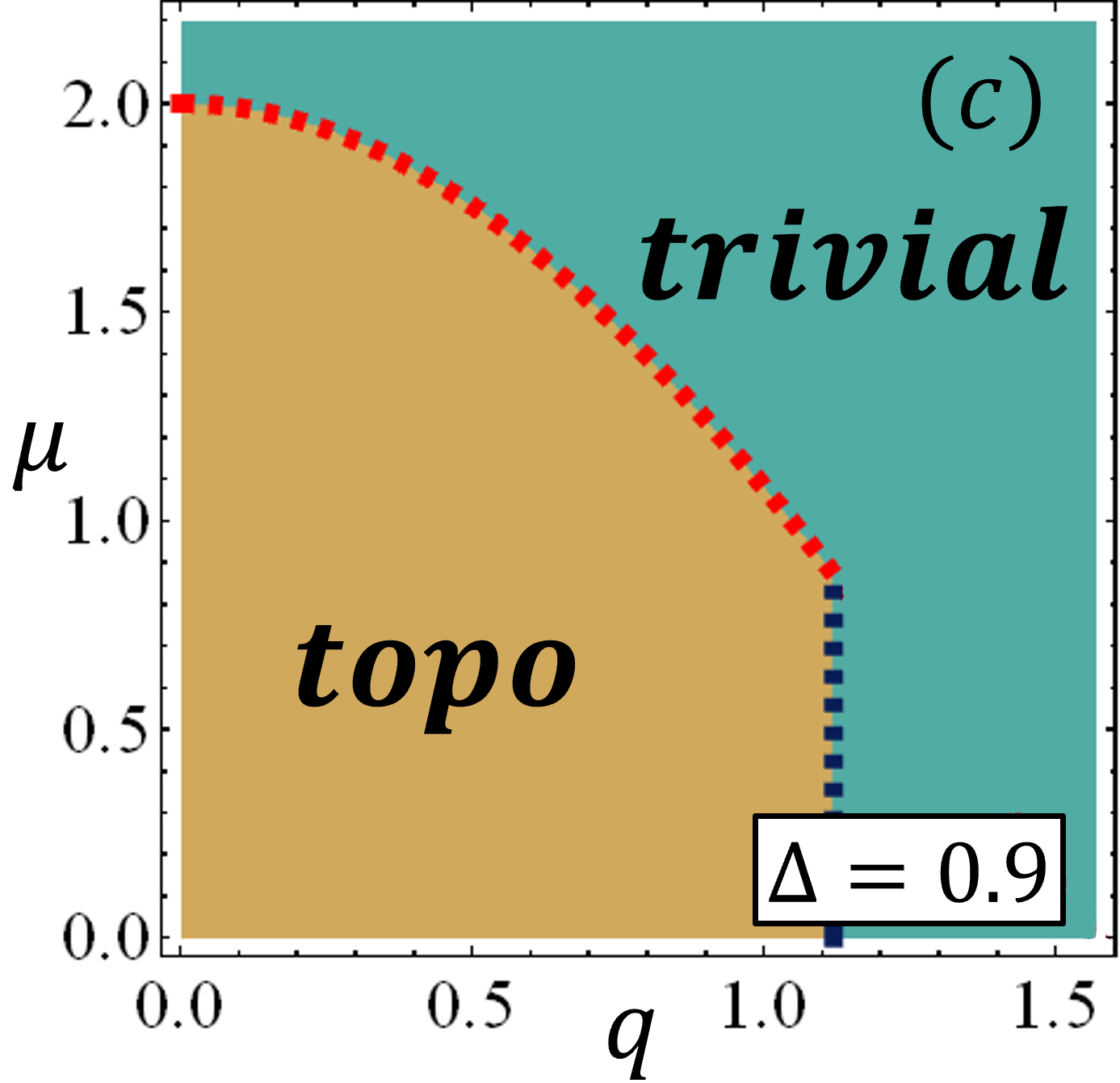}
	\includegraphics[scale=0.11]{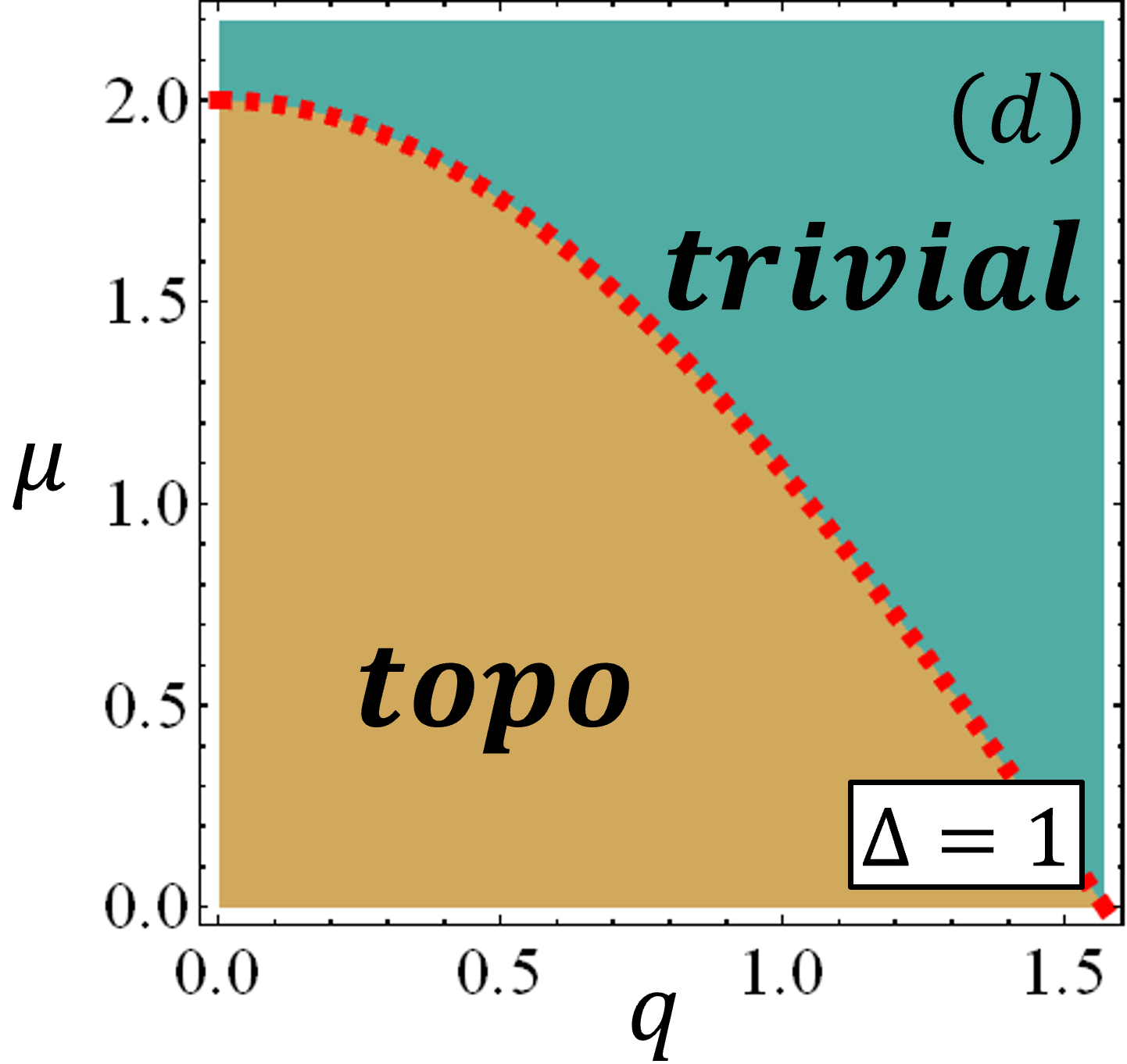}\\
	\vspace{0.2cm}
	\includegraphics[scale=0.09]{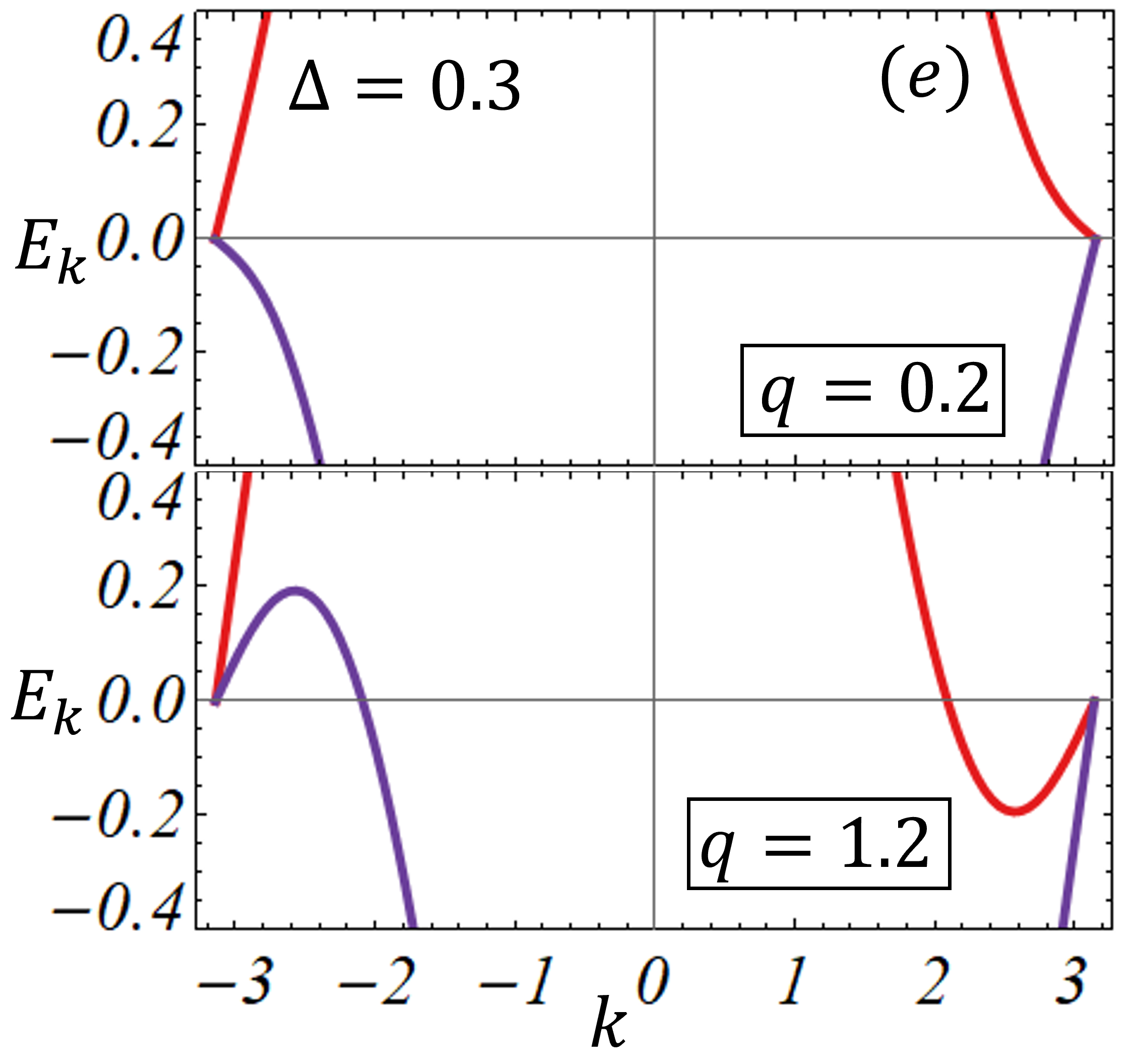}
	\includegraphics[scale=0.09]{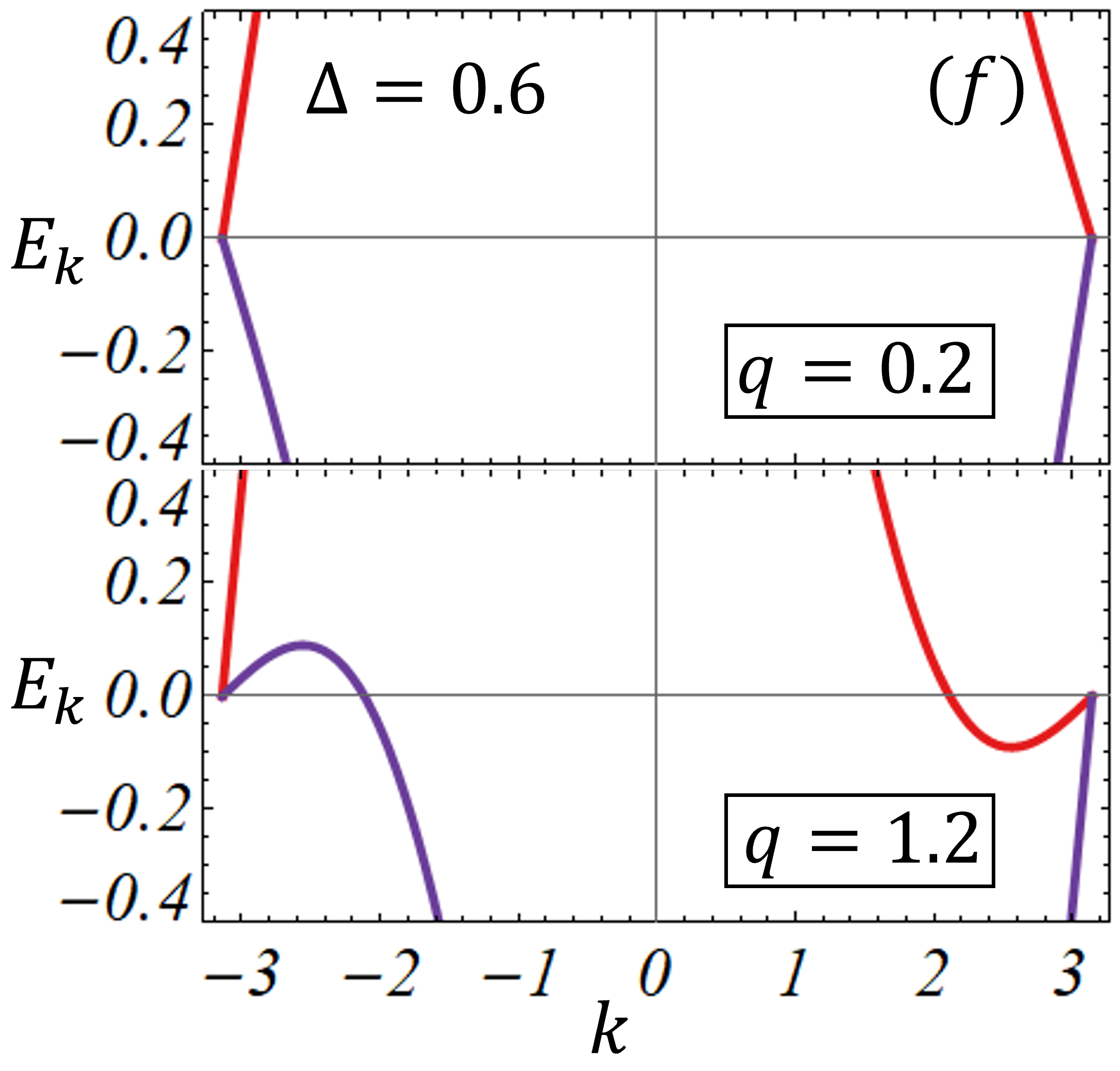}
	\includegraphics[scale=0.09]{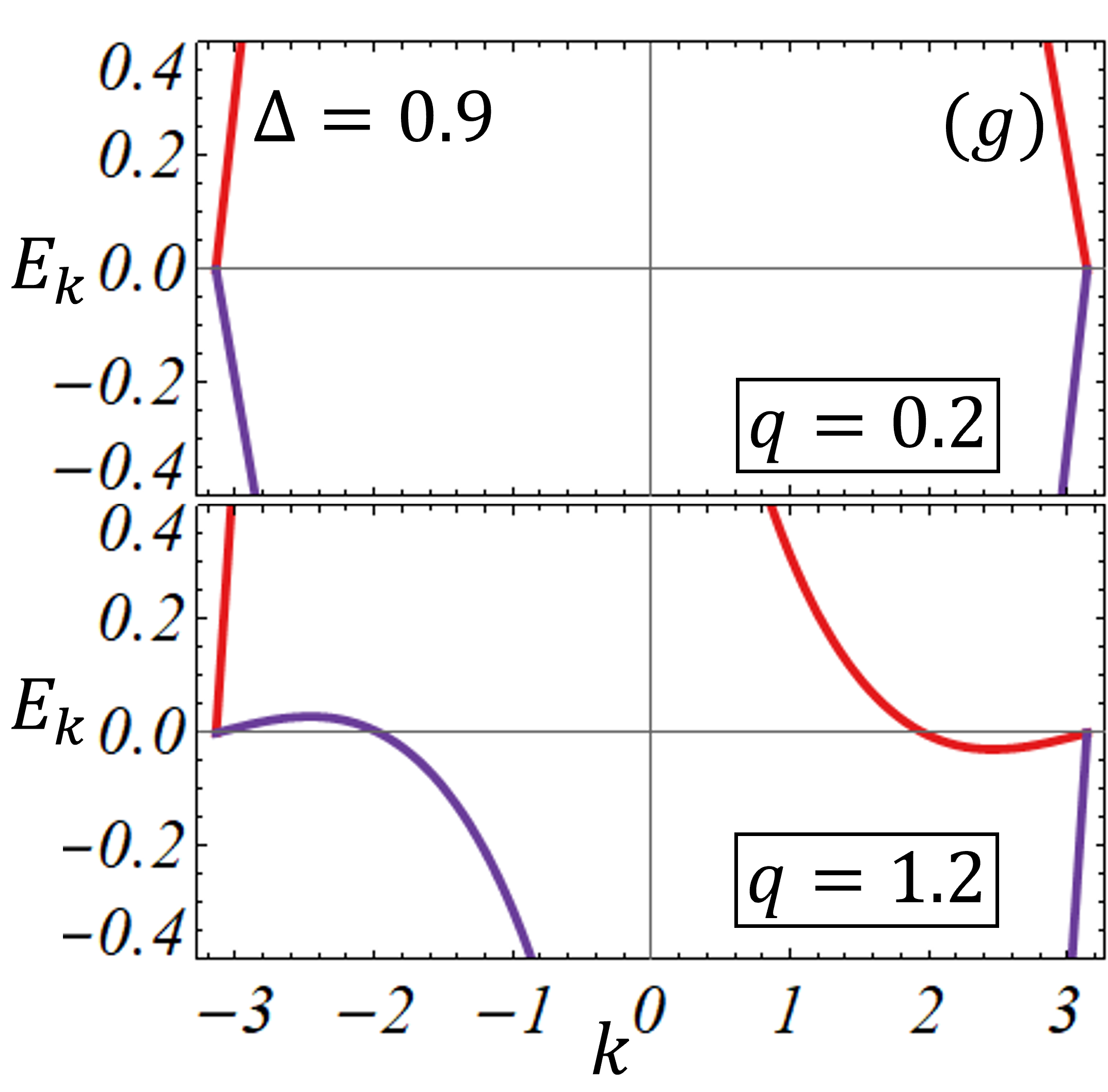}
	\includegraphics[scale=0.09]{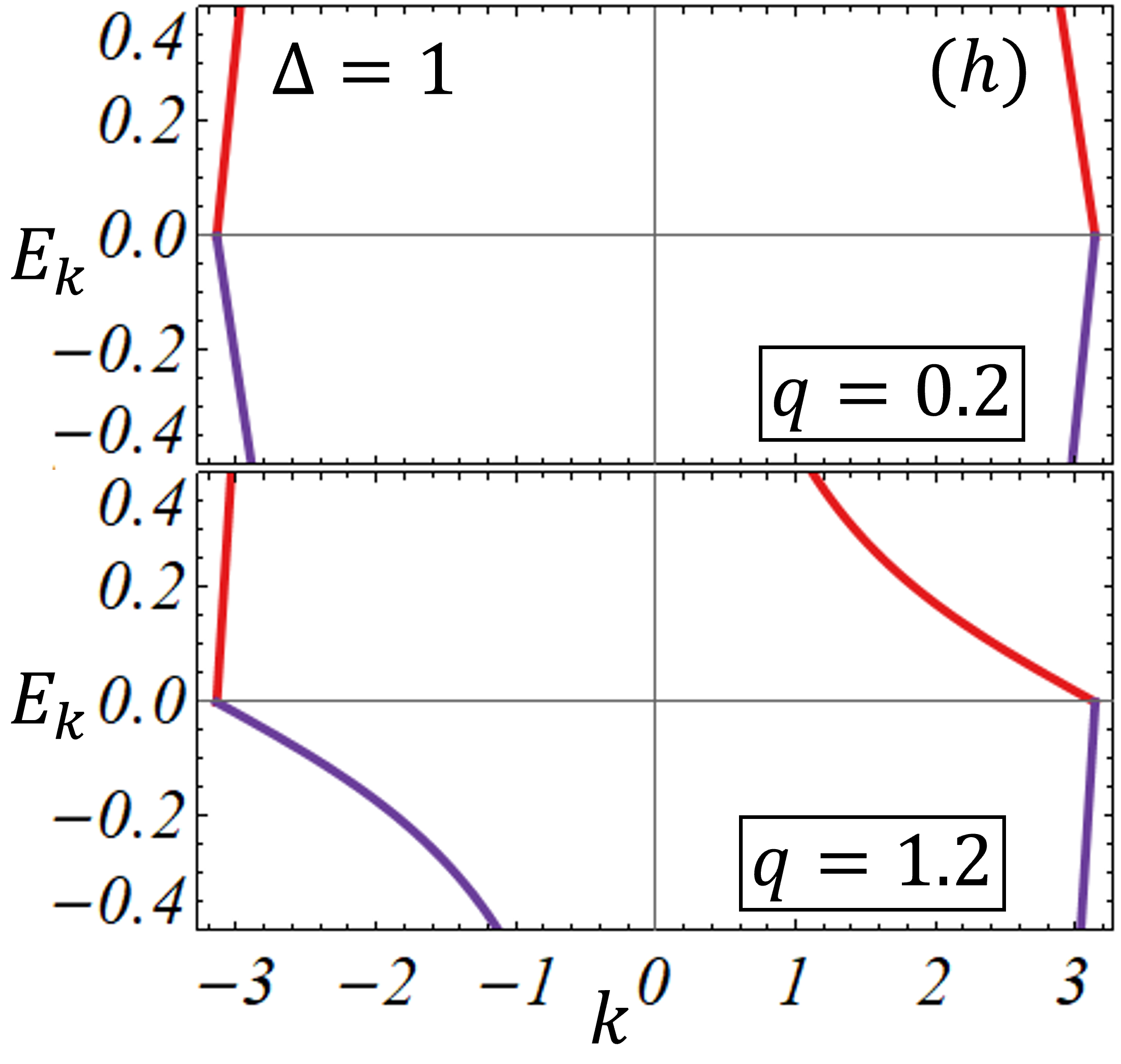}
	\caption{Bulk phase diagrams for $\Delta=0.3$ (a), $\Delta=0.6$ (b) , $\Delta=0.9$ (c), and $\Delta=1$ (d). Brown and cyan colours correspond respectively to topological and trivial phases. Red and blue curves, superimposed on phase diagrams, show the functions $\mu=2t\cos(q)$, $q=q_c=\arcsin(\Delta/t)$. The doubled panels (e)-(h) show the energy bands along the red curves of the phase diagrams (a)-(d), respectively at $q=0.2$ (top panels) and $q=1.2$ (bottom panels). The hopping strength is $t=1$ throughout.}
	\label{Fig1}
\end{figure*}

The paper is organized as follows. In Sec. \ref{bulk}, we introduce the model, study its main properties and discuss its bulk phase diagram. In Sec. \ref{RSmethods}, by means of real-space methods (Majorana polarization, edge-to-edge quantum conditional mutual information), we corroborate the bulk results by a systematic investigation of the conditions required to observe MBSs. Here we also discuss the special role played by the edge-to-edge quantum conditional mutual information, measuring the quantum correlations that arise between the system edges in the topologically ordered phase. In Sec. \ref{conclusions} we discuss our findings and possible future outlooks. Mathematical and technical details are presented and reviewed in Appendices \ref{AppA} and \ref{AppB}.

\section{Model, topology and bulk phase diagram}
\label{bulk}

\subsection{The Hamiltonian}

In the following we introduce an effective Hamiltonian for a Kitaev wire subject to the perturbing influence of a charged particle current. We start from the free Hamiltonian of a 
one-dimensional metal,
\begin{eqnarray}
H_f = - t\sum_{j=1}^{L-1} c^{\dagger}_j c_{j+1}+h.c. \, ,
\label{Hop}
\end{eqnarray}
whose band structure features the dispersion relation $\epsilon(k)= -2t \cos(k)$ in the thermodynamic limit. The group velocity of an electron with wave vector $k$ is then $v(k)=\partial_k \epsilon(k) / \hbar \sim 2t \sin(k)$. Since electronic states with $k$ and $-k$ are equally populated and $v(-k)=-v(k)$, no net current is observed in the system. 

This equilibrium picture breaks down when one considers a current flowing trough the system. The latter is a genuine nonequilibrium phenomenon that can be emulated by replacing the hopping strength $t$ in Eq. \ref{Hop} according to the prescription $t \rightarrow t e^{iq}$, where $q$ identifies the quasiparticle momentum induced by the current. The band structure of the modified free Hamiltonian is now shifted by the wave vector $q$ and reads $\epsilon(k)= -2t \cos(k-q)$. Accordingly, the group velocity takes the form $v(k) \sim 2t \sin(k-q)$, implying an average quasiparticle velocity proportional to $q$. 

Adding $p$-wave correlations on top of the metallic model, one ends up with a modified Kitaev chain Hamiltonian that includes the perturbing effect of a current flow:
\begin{eqnarray}
	H\!=\!\sum_{j=1}^{L-1} \biggl(-\!t e^ {iq} c^{\dagger}_j c_{j+1}\!+\!\Delta c_j c_{j+1}\!+\!h.c.\biggr)\!-\mu \!\sum_{j=1}^{L} c^{\dagger}_j c_j\!\, ,
	\label{KCcurrent}	
\end{eqnarray}
where the parameters $t$, $\Delta$, $\mu$ define, respectively, the nearest-neighbour hopping, the superconducting pairing and the on-site energy offset (chemical potential). The index $j \in \{1, \dots, L\}$ specifies the position along the lattice chain, while $c^{\dagger}_j$ and $c_j$ are the on-site fermionic creation and annihilation operators. 

Without loss of generality, the $q$-dependence can be moved to the $p$-wave pair potential by a $U(1)$ gauge transformation of the operators $c_j \rightarrow e^{-iqj} c_j$. As a result,  $\Delta \rightarrow \Delta e^{i2qj}$ and a Cooper pair acquires a finite momentum $2q$ \cite{gen66}, with $q$ a wave vector in the direction of the current flow. 

Within a condensed matter realization of Eq. \ref{KCcurrent}, it is expected that the current flow would give rise to rather small values of $q$, which is appropriate for the description of a d.c. current within the long wavelength limit. At any rate, in view of possible realizations via other quantum simulation platforms, for instance cold atomic gases, we will consider larger values of $q$ as well.

Finally, the Bogoliubov-de Gennes representation of Eq. \ref{KCcurrent} can be obtained by introducing the Nambu spinors in momentum representation: $\Psi(k)=(c_k,c^\dagger_{-k})^T$, so that we obtain $H=1/2 \sum_{k}\Psi^\dagger(k)\tilde{H}(k)\Psi(k)$, with
\begin{eqnarray}
	\label{BdGKspace}
	\tilde{H}(k)=\left(
	\begin{array}{cc}
		-2t \cos(k-q)-\mu&2i \Delta \sin(k)\\
		-2i \Delta \sin(k)&2t \cos(k+q)+\mu\\
	\end{array}
	\right).
\end{eqnarray}\\

\subsection{Topology and bulk phase diagram}

In equilibrium conditions, topological phases of mean field Hamiltonians are meaningfully described by bulk topological invariants, according to the Altland-Zinbauer ten-fold classification \cite{PhysRevB.55.1142}. The ten symmetry classes \cite{PhysRevB.55.1142} allow to associate the appropriate topological invariants to the bulk Hamiltonians according to the dimensionality and the simultaneous presence/absence of particle-hole symmetry ($P$), time reversal symmetry ($T$) and chiral symmetry ($C$). The topological invariants capture the topology (in mathematical sense) of the band structure of the bulk, providing the phase diagrams of the systems in thermodynamic limit, also identifying the band gap closing points. 

The original Kitaev model, whose Hamiltonian is obtained by setting $q=0$ in Eq. \ref{BdGKspace}, belongs to the BDI class of the ten-fold classification since it enjoys all the three aforementioned symmetries. 
In the presence of a flux, $q \neq0$, the time reversal symmetry and the chiral symmetry ($C=P T$) break down, leaving the particle-hole as the only protecting symmetry of the topological phase. Thus, the charge current leads the chain from BDI class with $\mathcal{Z}$ index to D class with $\mathcal{Z}_2$ index of the Cartan classification. 

The time-reversal symmetry breaking mechanism induced by the current has significant implications in the relationship  between topology of isolated systems and measurement procedures. Indeed, several time-reversal protected systems, belonging to classes BDI,CI, CII, can host multimode phases where more then a single non-trivial mode nucleates at the edge of the system \cite{Zhou_2016,Maiellaro2018TopologicalPD,YAN}. On the other hand, breaking $T$ by adding a particle current reduces the total number of symmetries to at most one and simultaneously induces a change of class of the ten-fold classification. This mechanism implies that in some cases one single edge mode holds robust to the measurement procedure, while the other ones are fragile against the injected current. This can be the case when a one-dimensional multi-leg Kitaev ladder in the BDI class collapses into the D class. 

The connection between symmetries and topology can be formalized in a rigorous manner and the topological invariant $Q$ can be formally defined even when $q \neq 0$, as discussed in Appendix \ref{AppA}. Indeed, the $Q$ parameter can be deduced by looking at the band properties, i.e. by identifying the bulk gap closing points. Hence, by looking at the analytical expression of the energy bands $E_1(k)$ and $E_2(k)$, gap closing points are obtained as real solutions of the equations $\mu=-2t \cos(k) \cos(q)\pm \sqrt{\phi_c \sin(k)^2}$, with $\phi_c=t^2-2 \Delta^2-t^2 \cos(2q)$. When $q< \arcsin(\Delta/t)$, gap closes only at $k=0$ or $\pi$, corresponding to phase boundaries $\mu=\pm 2t \cos(q)$, respectively (Fig. \ref{Fig1} upper panels (e)-(h)). For $q\geq \arcsin(\Delta/t)$ the system only shows band crossing points and, as a consequence, it is expected to be in a trivial phase (Fig. \ref{Fig1} lower panels (e)-(h)). 

The above discussion hints that the system experiences a topological phase transition at the boundary of the plane region defined by $|\mu| < 2 t \cos(q) \land q < \arcsin(\Delta/t)$. This criterion leads to the topological invariant $Q$ defined in Appendix \ref{AppA} and to the phase diagrams reported in Fig. \ref{Fig1}, panels (a)-(d), where the curves $\mu=2t \cos(q)$ (red) and $q=q_c=\arcsin(\Delta/t)$ (blue) partially overlap with the boundaries of the topological phase. As shown in panels (a)-(d) of Fig. \ref{Fig1}, a critical value $q_c$  exists for which the superconducting order and the topological regime are simultaneously lost. Actually, the existence of such limit is expected in a superfluid and it is reminiscent of the Landau critical velocity \cite{ZagoskinBook}. Below this threshold, MBSs are resilient and the phase boundary features only a $q$-dependent renormalization which is well approximated by $\mu_c \sim 2t (1-q^2/2)$. The latter observations provide a direct proof of the resilience of topological order against a moderate amount of current injected into the system. 

In order to further validate this physical picture, in the following we identify the phase diagram according to two different nonlocal indicators of topological order, namely the Majorana polarization and the nonlocal correlation that is established between the edges as quantified by the quantum conditional mutual information.

\section{Nonlocal topological order parameters}
\label{RSmethods}

\begin{figure*}
	\includegraphics[scale=0.14]{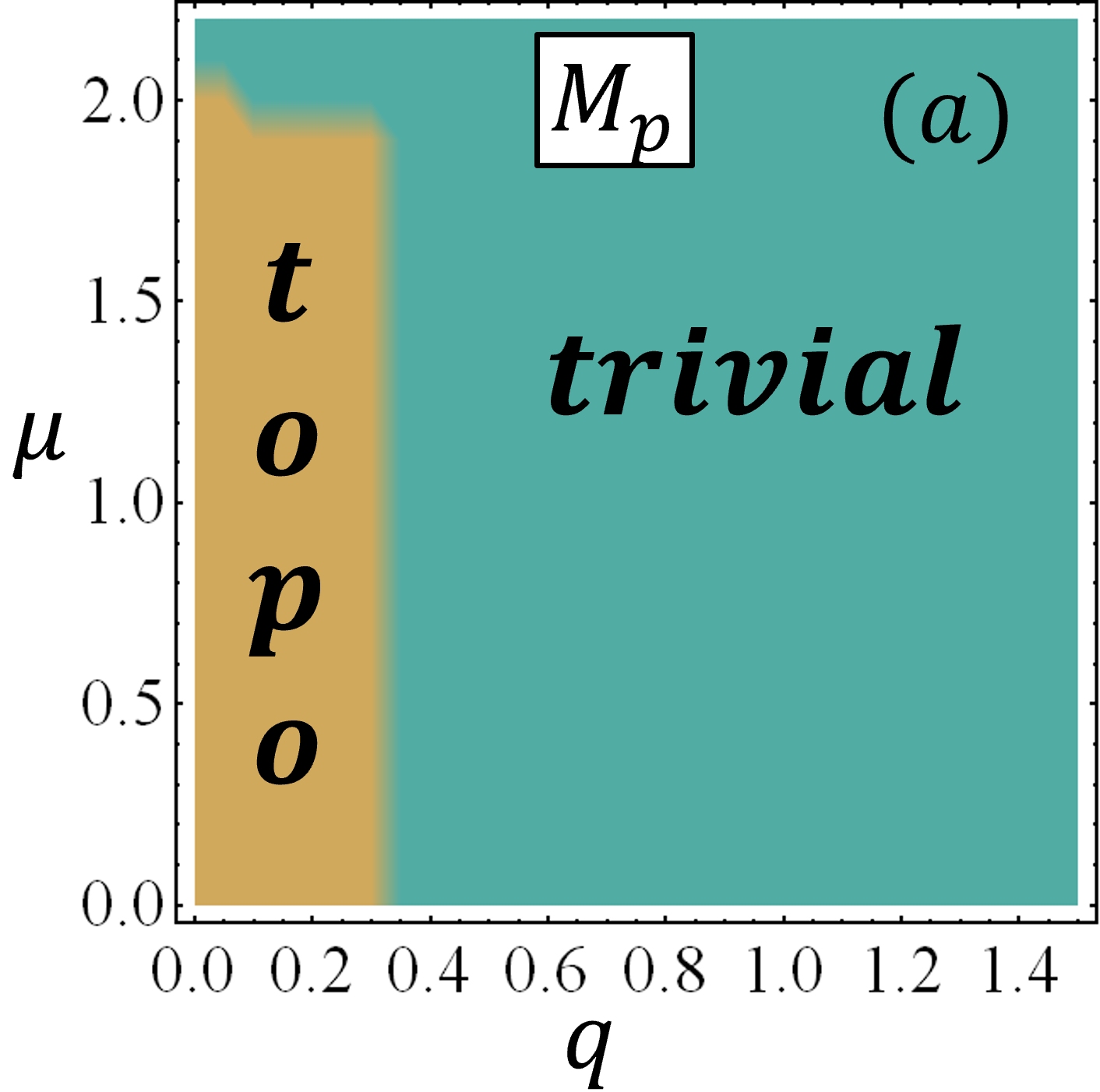}
	\includegraphics[scale=0.14]{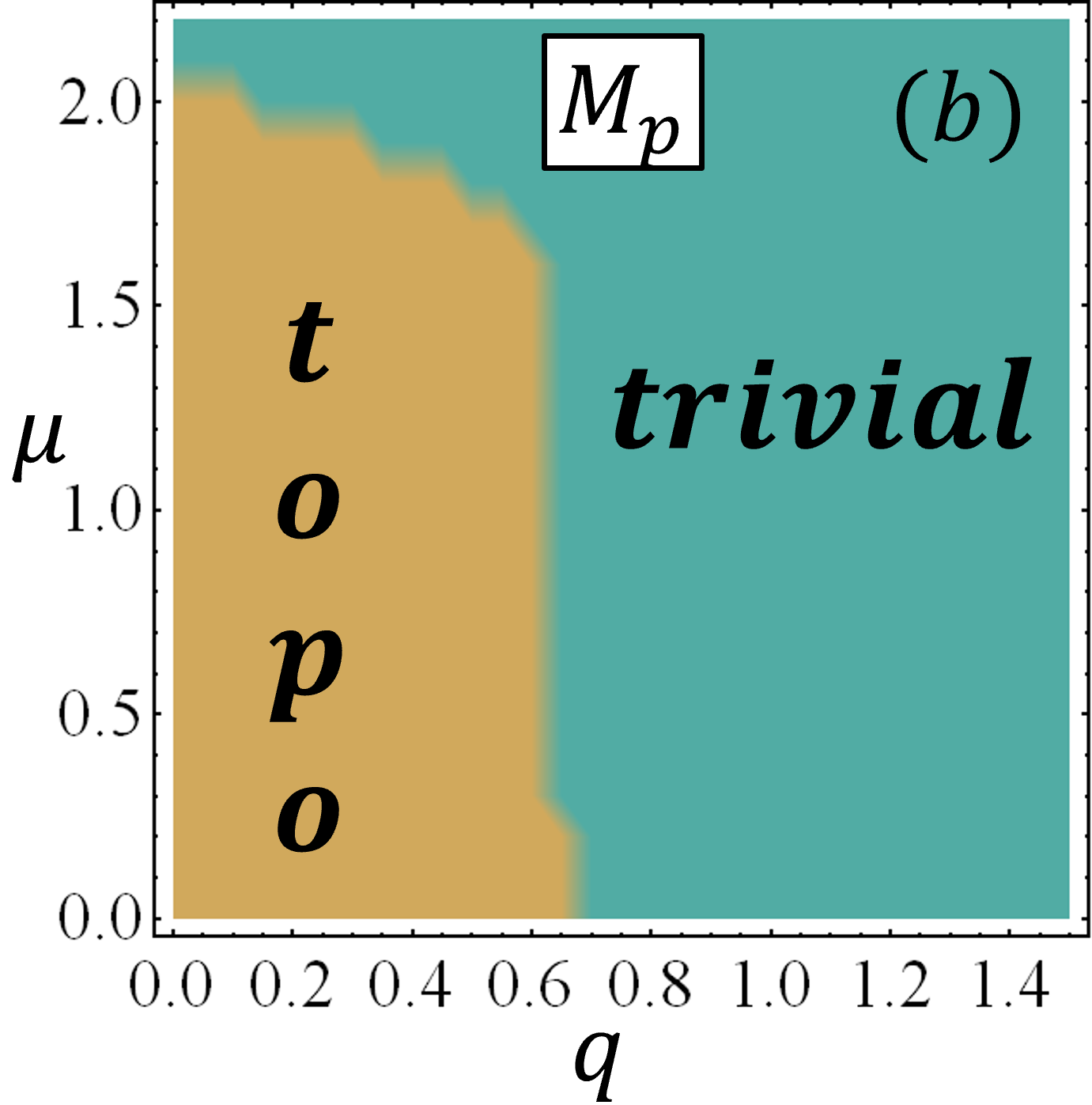}
	\includegraphics[scale=0.14]{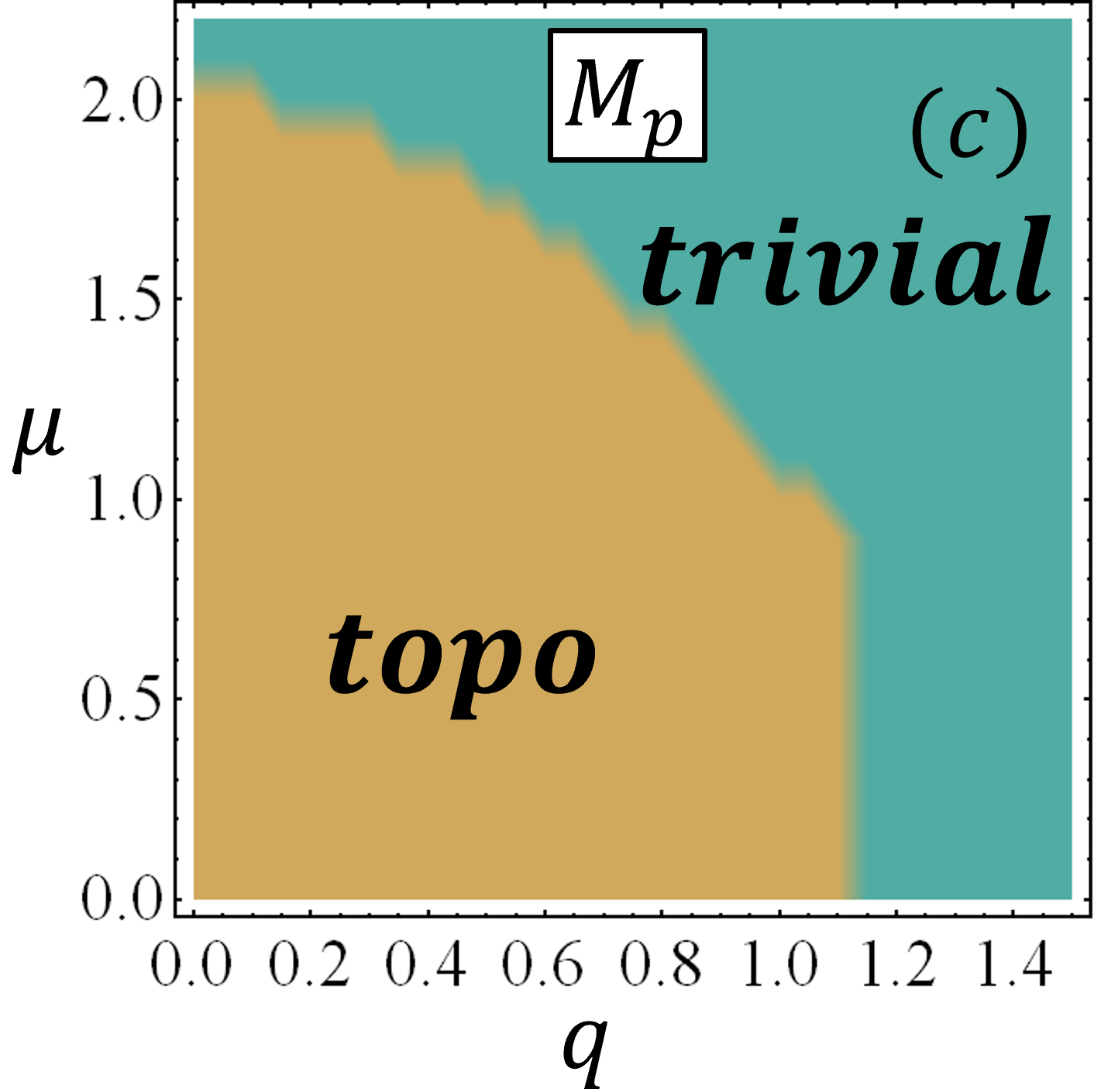}
	\includegraphics[scale=0.14]{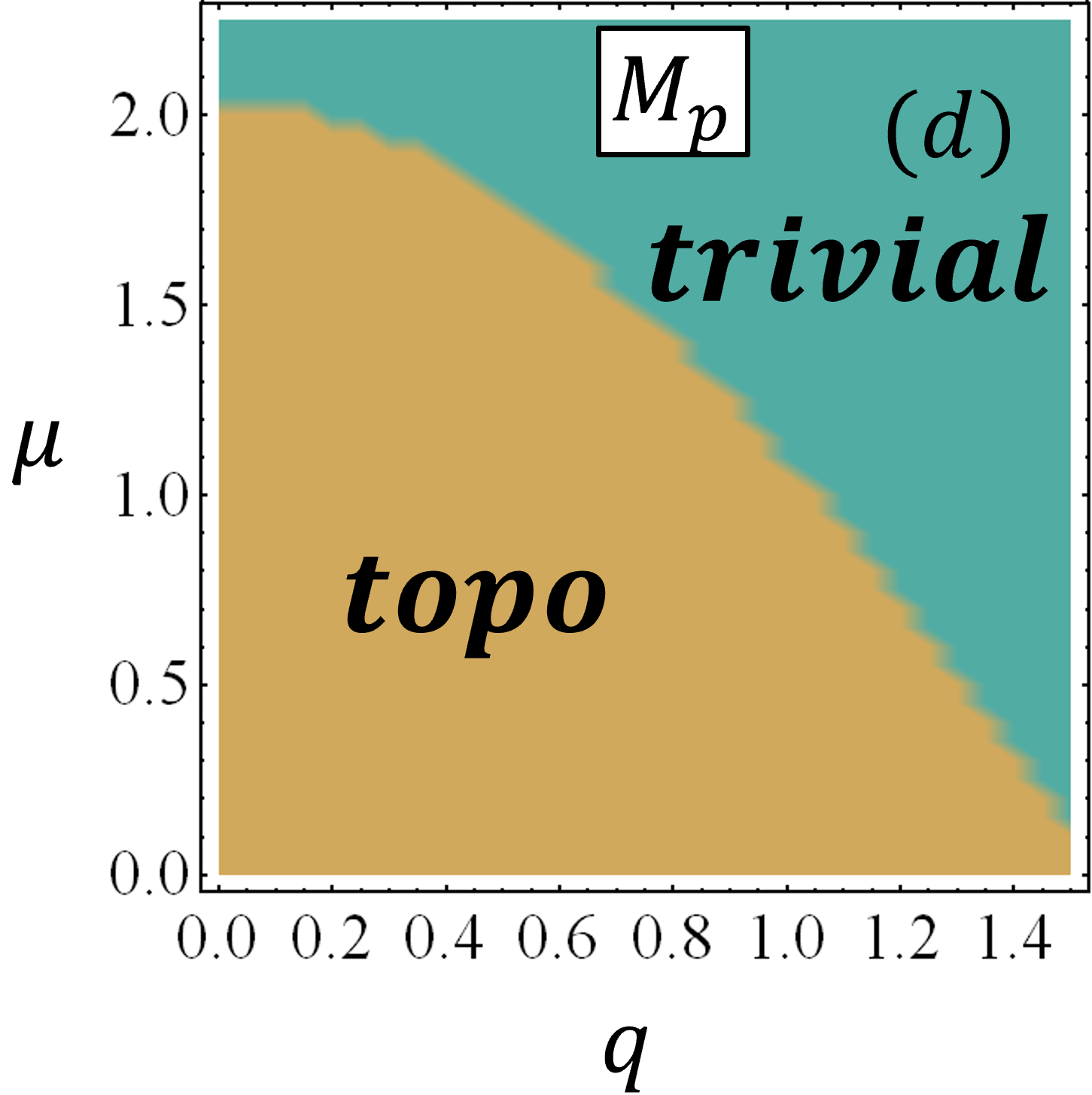}\\
	\vspace{0.2cm}
	\includegraphics[scale=0.14]{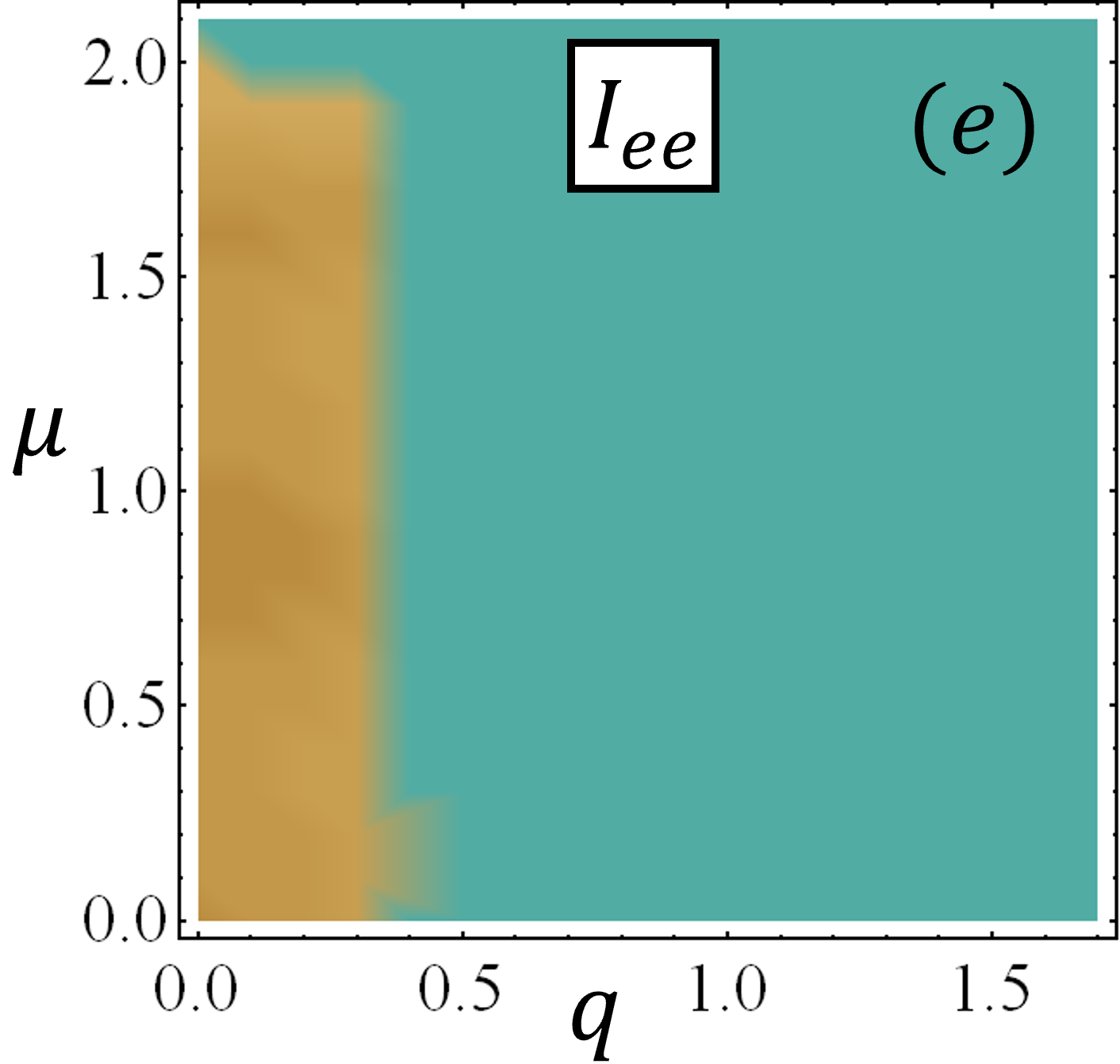}
	\includegraphics[scale=0.14]{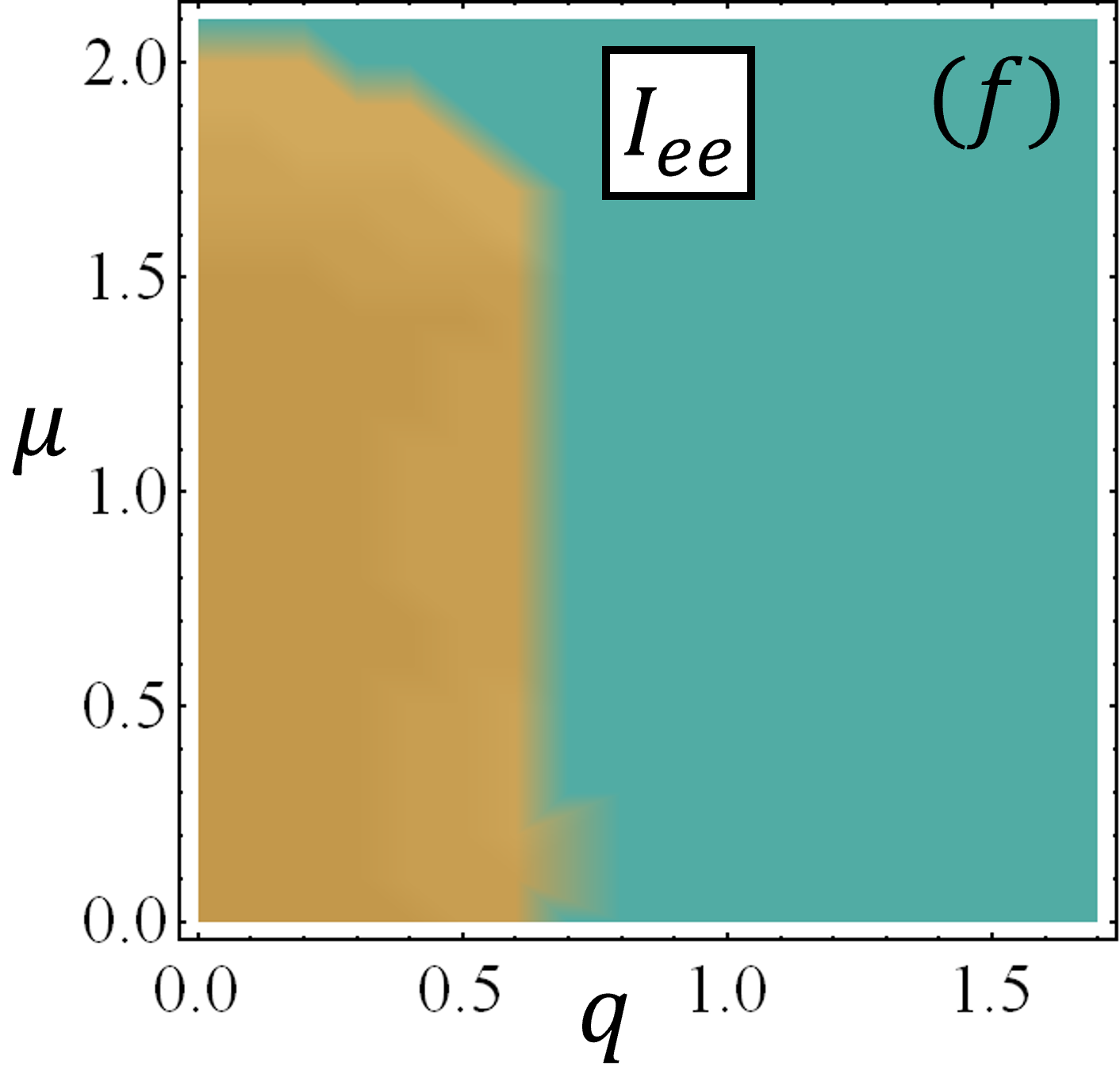}
	\includegraphics[scale=0.14]{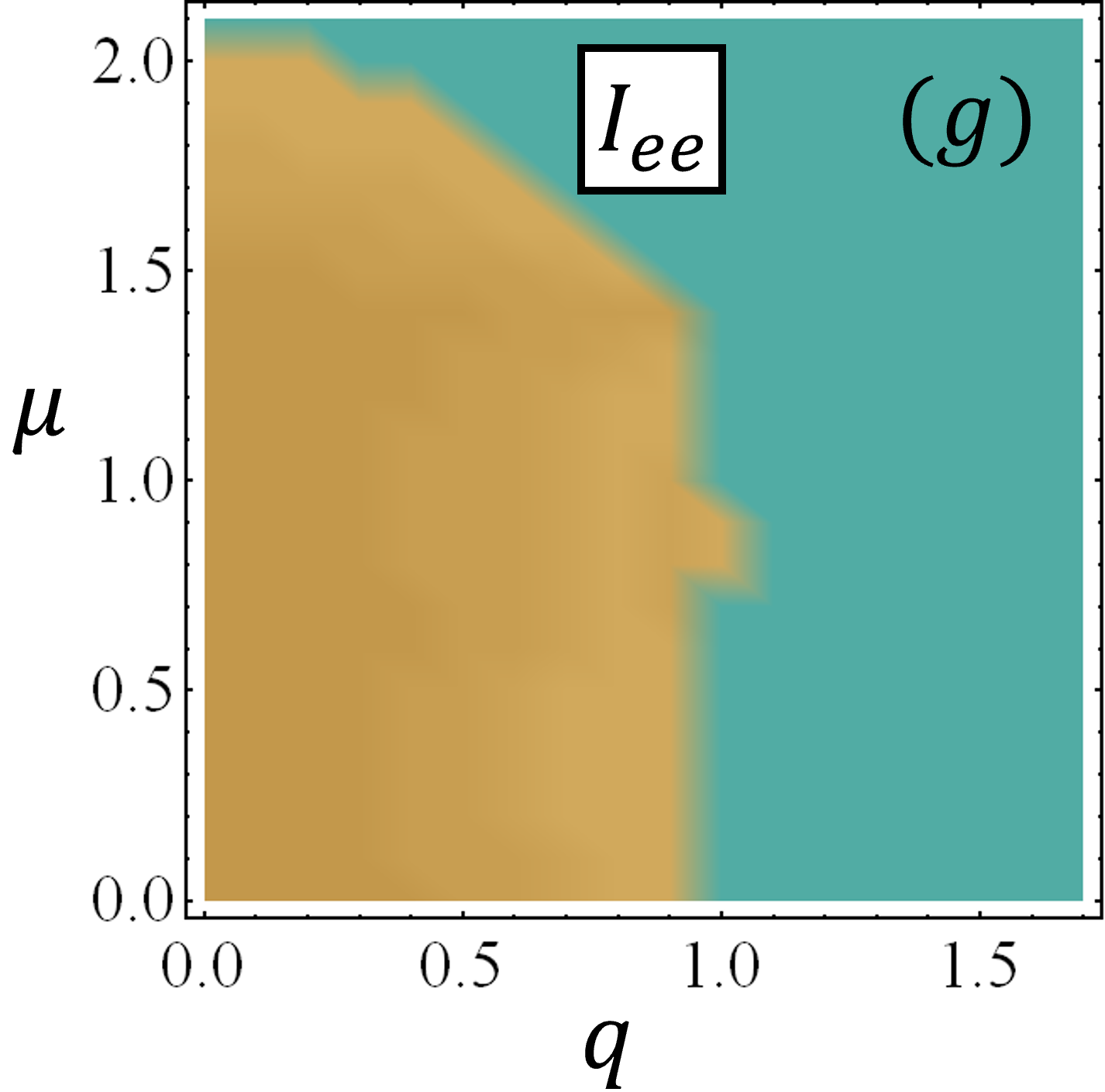}
	\includegraphics[scale=0.14]{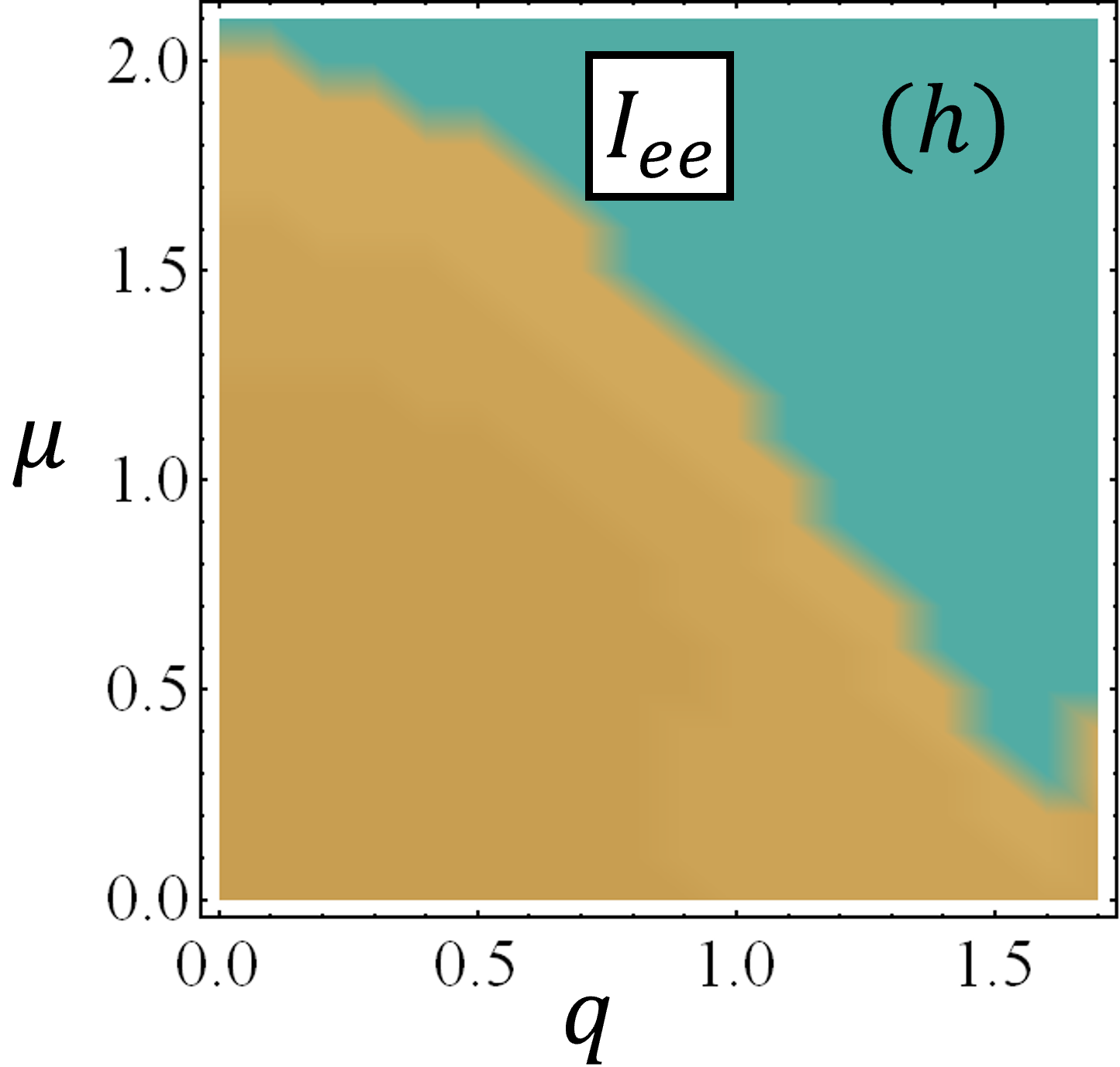}
	\caption{Real-space phase diagrams according to the Majorana polarization $M_p$, panels (a)-(d), and according to the QCMI $I_{ee}$, panels (e)-(h), for the same parameters choice adopted in Fig. \ref{Fig1}. One has $M_{p}=1$, $0$ and $I = \log2/2$, $0$ respectively in the topological and in the trivial phase. The size of the system is set at $L=100$ in panels (a)-(d). In panels (e)-(h) the size of the system and the size of the edges are, respectively, $L=10$ and $L_A=L_B=3$.}
	\label{Fig2}
\end{figure*}

Real-space nonlocal order parameters, such as the Majorana polarization (MP) \cite{BENA2017349,PhysRevLett.110.087001,PhysRevB.85.235307,MaiellaroGeoFrust} and edge-to-edge quantum conditional mutual information (QCMI) \cite{PhysRevResearch.4.033088,MaieIllum2,PhysRevB.106.155407}, have been proposed and extensively used to investigate the presence/robustness of topological, symmetry-protected edge states. The MP and the QCMI capture complementary aspects of MBSs. More specifically, MP measures the weight of the Majorana quasiparticles in Nambu space. Following the notation of Refs. \cite{BENA2017349,PhysRevLett.110.087001,PhysRevB.85.235307,MaiellaroGeoFrust}, the MP can be expressed as follows:
\begin{equation}
M_p(j,\omega)=\sum_{n}\bigl( u_{n,j} v_{n,j}\bigr) (\delta(\omega-E_n)+\delta(\omega+E_n)) \, , 
\end{equation}
with $u$ and $v$, the particle and hole weights in Nambu representation. In particular, by choosing $\omega=0$, the total MP $M_{p}=|\sum_{j=1}^{L/2} M_p(j,0)|$ is equal to $1$ for genuine MBSs, vanishes for electrons/holes, and decreases from the maximum value $1$ for hybridized modes originated by genuine initial MBSs.

On the other hand, the edge-to-edge QCMI $I_{ee}$ determines the unique, long-distance and nonlocal quantum correlations that are established in a topologically ordered phase between the system edges. Indeed, such topological nonlocal edge-to-edge correlations are faithfully quantified by a specific measure of bipartite entanglement \cite{doi:10.1063/1.1643788,5075874}, the squashed entanglement (SE) $E_{SQ}^0$ between the edges. Taking a tripartition of a one-dimensional system in terms of edge $A$, edge $B$, and bulk $C$, the SE between A and B is defined as the minimum of the QCMI between $A$ and $B$ taken over all possible $C$-extensions of the system, keeping $A$ and $B$ fixed \cite{PhysRevResearch.4.033088,MaieIllum2}. The edge-edge QCMI $I_{ee}$ thus provides the natural quantum upper bound on the true long-distance SE between the edges. It is defined by a suitable combination of the reduced von Neumann entropies between the connected and disconnected parts of the tripartite system, namely:
\begin{equation}
I_{ee} = S_{AC} + S_{BC} - S_{C} - S_{ABC} \, .
\label{QCMI}
\end{equation}
The first three terms in the rhs  of Eq. (\ref{QCMI}) are the von Neumann entropies of the ground-state reduced density matrices, respectively for subsystems $AC$ (left edge and bulk, after tracing out the right edge), $BC$ (right edge and bulk, after tracing our the left edge), and $C$ (bulk, after tracing out both edges). The last term is the total ground-state von Neumann entropy that vanishes whenever the ground state is a pure state. The particular combination of total and reduced entropies in Eq. (\ref{QCMI}) "squashes" out the classical contributions to the total correlations, leaving only the genuine quantum contributions to the correlations between the edges \cite{PhysRevResearch.4.033088}.  

\begin{figure*}
	\includegraphics[scale=0.09]{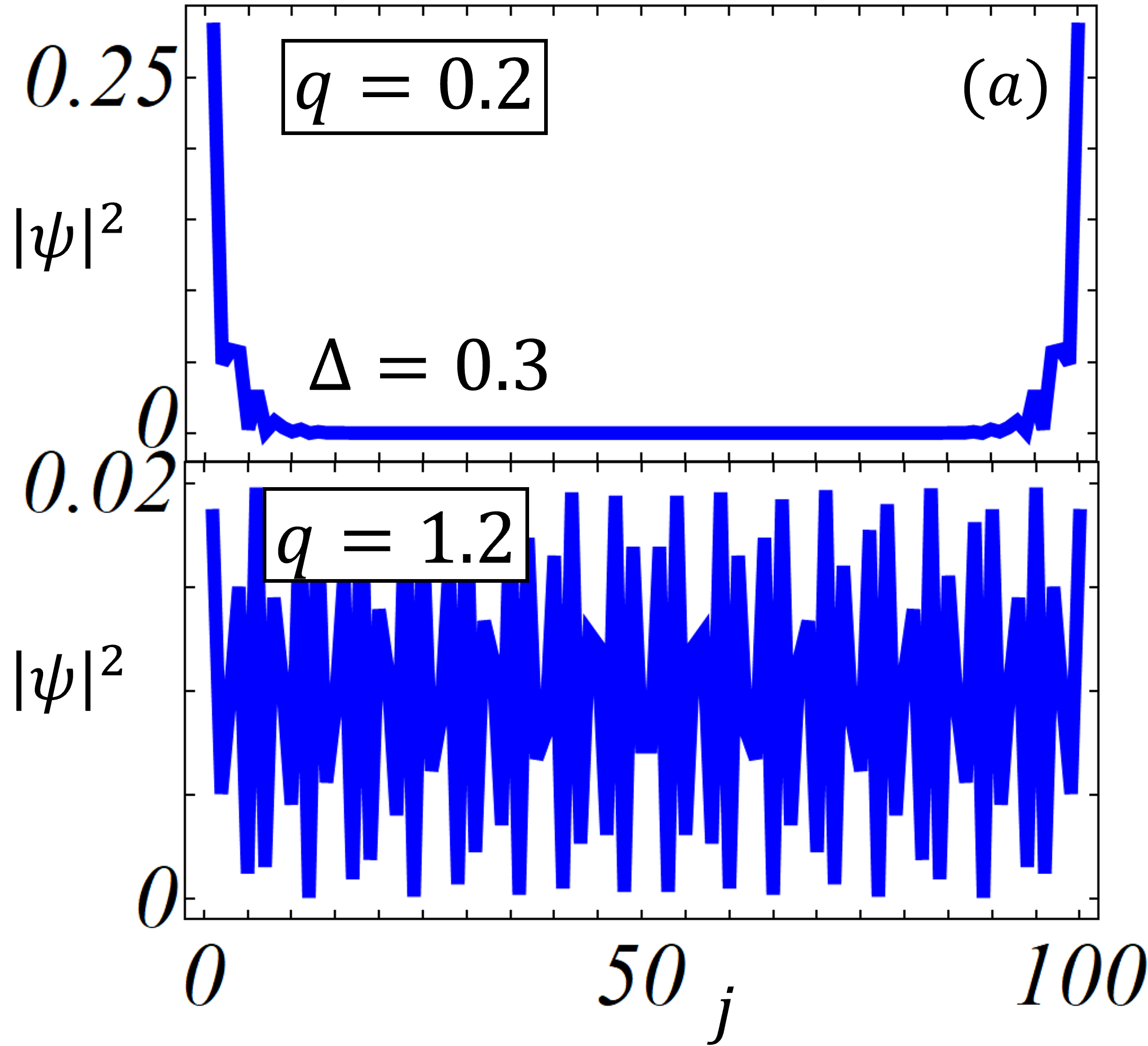}
	\includegraphics[scale=0.09]{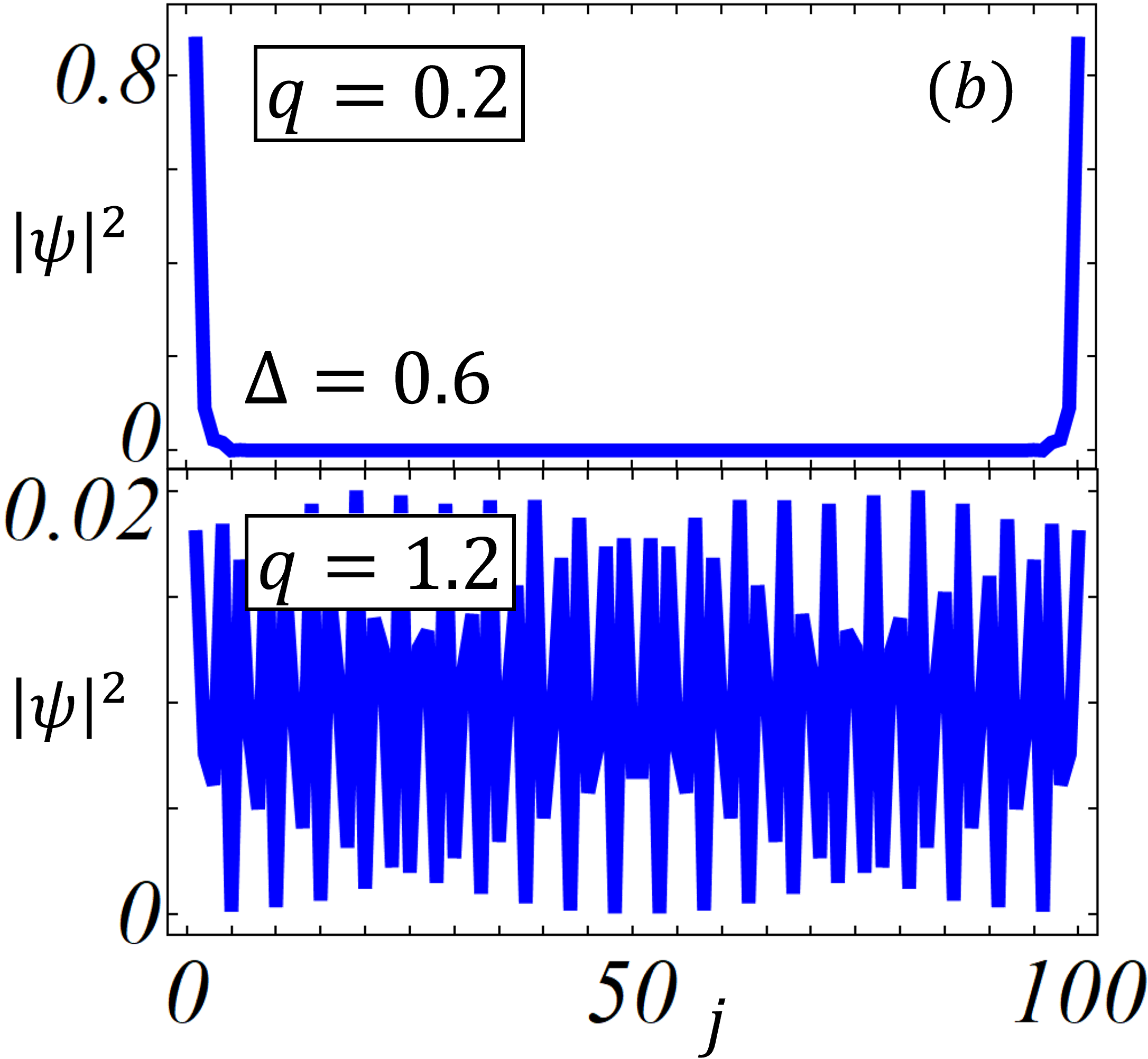}
	\includegraphics[scale=0.09]{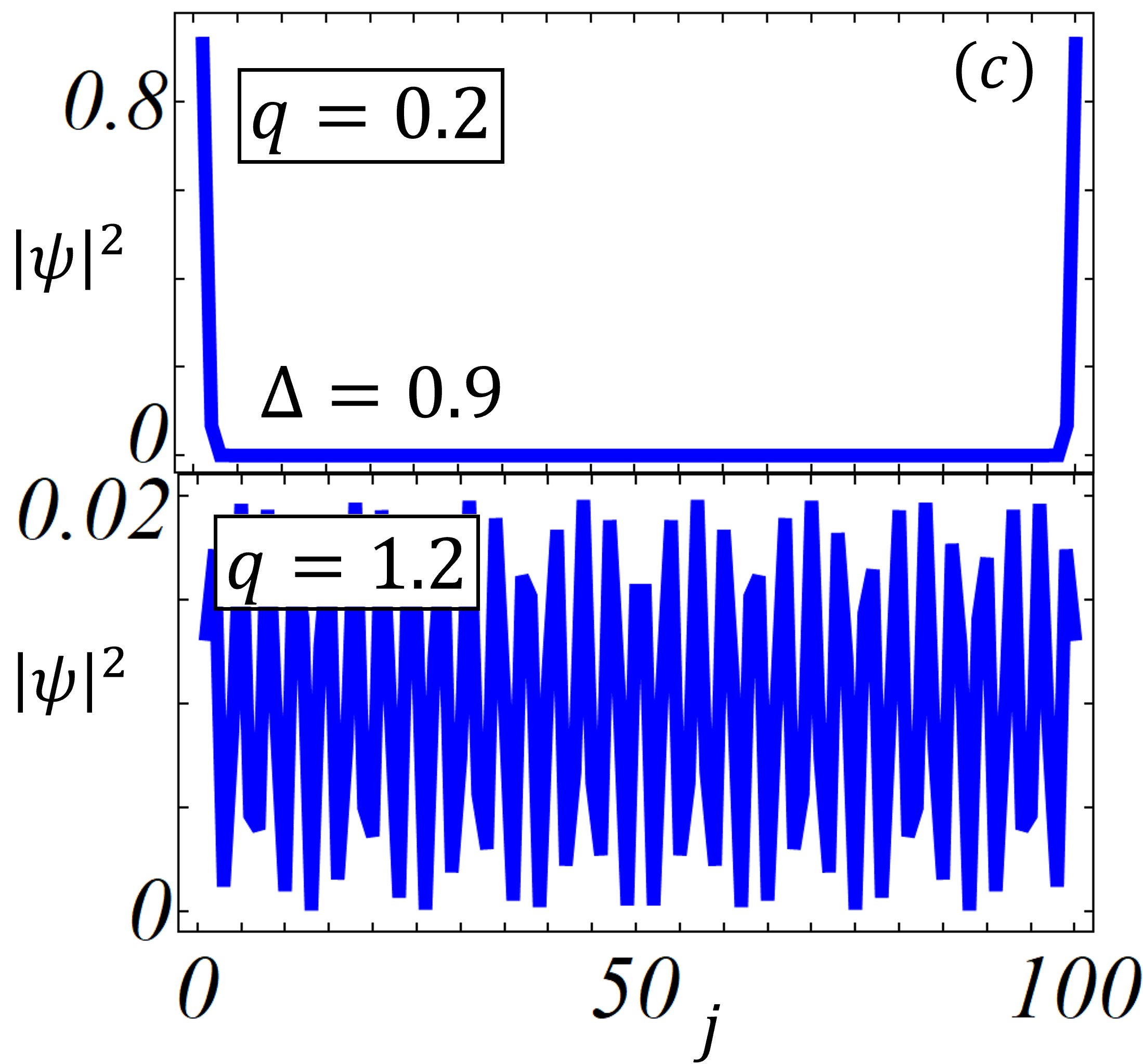}
	\includegraphics[scale=0.09]{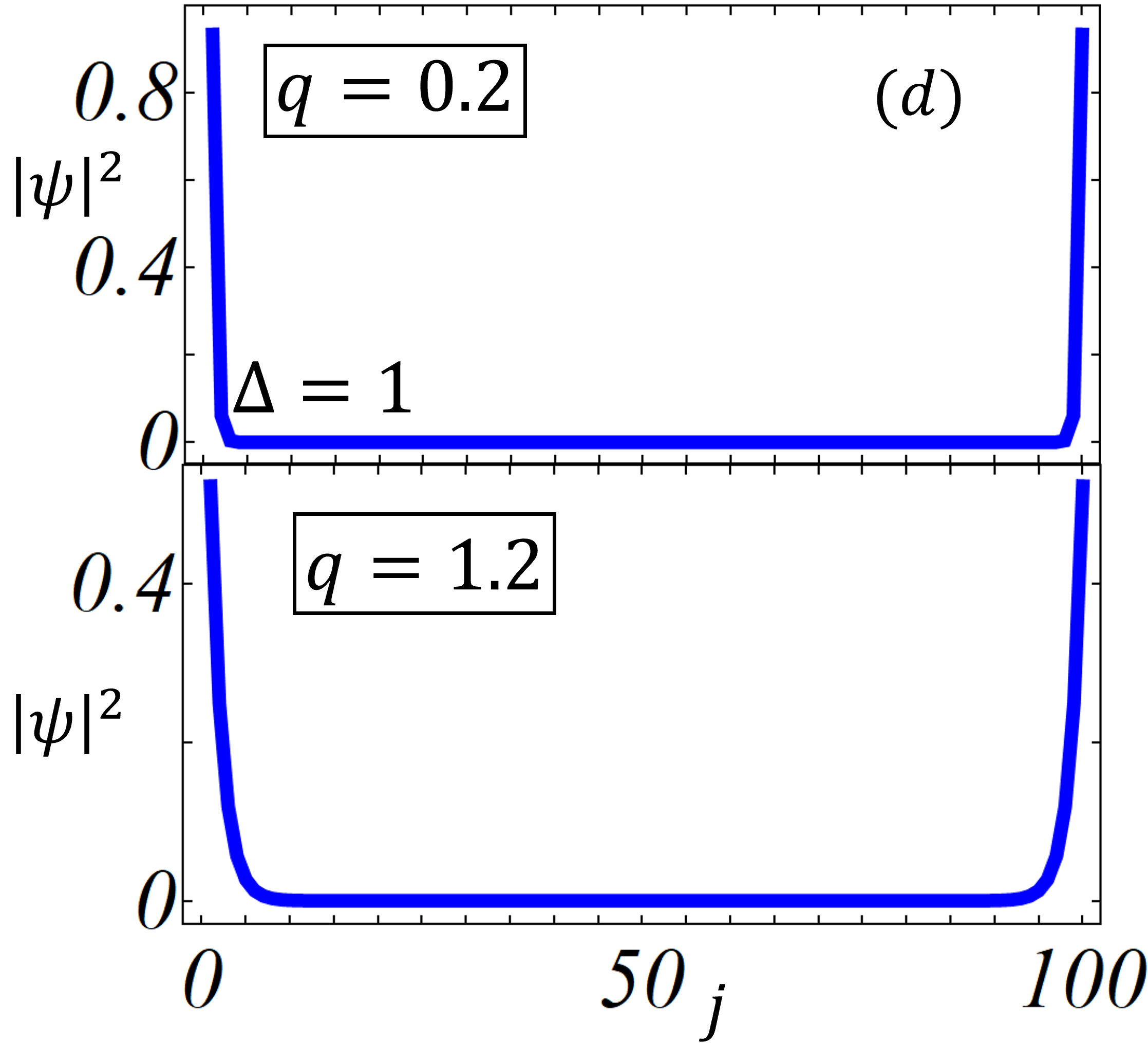}
	\caption{Modulus squared of the lowest energy eigenstates in agreement with the phase diagrams of Fig. \ref{Fig1}. We also fix $\mu=0.5$ and $L=100$. MBSs and trivial states are clearly recognised for $q=0.2$ and $q=1.2$ in panels from (a) to (c). MBSs appear for both $q=0.2$ and $q=1.2$ in panel (d).}
	\label{Fig3}
\end{figure*}

As symmetry-protected topological order is encoded in the edges, the edge-edge QCMI $I_{ee}$ identifies unequivocally topologically ordered phases, satisfying all the criteria of a genuine nonlocal order parameter. In particular, the edge-edge QCMI takes the quantized value $I_{ee} = \log{2}/2$, i.e. half of the maximal Bell-pair entanglement, at the exact ground-state topological degeneracy point, $\mu = 0$, for a Kitaev chain with open boundary conditions hosting genuine Majorana modes, and remains constant at this quantized value throughout the entire topological phase, i.e. up to $\mu = 2t$ \cite{PhysRevResearch.4.033088}. 

Such behavior is actually typical in one-dimensional topological quantum matter. For instance, one finds that $I_{ee} = \log{2}$ throughout the topologically ordered phase of the SSH topological insulator \cite{MaieIllum2}. This is exactly the maximal Bell-state entanglement, as should be expected for a system whose edge modes are standard Dirac fermions (topological insulator) and not "half-fermion" Majoranas (topological superconductor). These analytical coincidences on different classes of topological systems lead to conjecture that the QCMI nonlocal topological order parameter $I_{ee}$ is not only an upper bound on the true edge-edge squashed entanglement $E_{SQ}^0$, but in fact coincides with it in the ground state of all one-dimensional symmetry-protected topological systems \cite{PhysRevResearch.4.033088,MaieIllum2}.

Resorting to the Jordan-Wigner mapping \cite{Franchini2017}, one can transform the fermionic degrees of freedom into spins distributed along a one-dimensional lattice. The resulting model Hamiltonian is that of an $XY$ spin chain, modified by a mixing term between the $X$ and $Y$ components of the spins (see Appendix \ref{AppB} for details). The mapping allows several computational advantages in the evaluation of the various reduced von Neumann entropies, either connected or disconnected.

The phase diagrams in Fig. \ref{Fig2}, obtained by means of the MP $M_{p}$, panels (a)-(d), and by means of the edge-edge QCMI $I_{ee}$, panels (e)-(h), provide a consistent picture of the topological phases of the system and are consistent with the bulk phase diagrams reported in Fig. \ref{Fig1}. The correspondence between these three types of phase diagrams confirms that MBSs survive to a moderate amount of current flow and proves that the bulk-edge correspondence is an intrinsic and meaningful property of topological materials even in nonequilibrium conditions. 

Due to finite-size effects, the few modest quantitative discrepancies are observed at the phase boundaries. In fact, when using the MP $M_p$ we can afford setting a system size $L=100$, while when resorting to the QCMI $I_{ee}$ we set the system size at only $L=10$. This difference is due to the different computational resources needed to evaluate the two quantities; indeed, calculating $I_{ee}$, a much more sophisticated quantity, involves keeping track of all different reduced states, with the respective eigenvalues and eigenvectors, through all the different subsystem partitions of increasing size.
The excellent qualitative agreement between the two phase diagrams despite a difference of one order of magnitude in the system size suggests that non-trivial long-distance correlations between MBSs are more robust to finite-size effects than their spectral properties. 

Finally, the spatial profile of the lowest energy modes that we report in  Fig. \ref{Fig3} is also consistent with the emerging picture. In particular, we observe localized modes at the edges with a decaying tail in the bulk for values of the Hamiltonian parameters corresponding to the topologically ordered phase, while completely delocalized modes are observed in trivial phase.

\section{discussion and outlook}
\label{conclusions}

In conclusion, we have studied the topological properties of a Kitaev chain under the perturbing influence of a uniform charged current injected into the system. This investigation sheds light on the stability of the topological phases of open systems subject to measuring processes. We have proved the robustness of topological phases under a moderate current. Indeed, when the current flow exceeds a critical threshold, superconducting correlations and topological order are simultaneously lost. On the other hand, below such threshold the edge modes turn out to be robust, even though the extension of the topological phase is reduced compared to the case of an unperturbed Kitaev chain. We have also shown that the current induces a time-reversal symmetry breaking and reduces the number of protecting symmetries of the chain. The latter is a rather general mechanism that reveals the fragility of some classes of topological materials to measurement procedures. Indeed, when the current is applied to one-dimensional BDI systems hosting more than a single edge mode, due to the symmetry reduction mechanism, most of the modes are destabilized, while at most one single mode remains stable against the injected current.

We have investigated the resilience of topological states by using several physical indicators, including the Majorana polarization and the recently introduced edge-edge quantum conditional mutual information $I_{ee}$ that provides crucial information about the nonlocal quantum correlations shared by the edge Majorana excitations. These real-space methods, complemented by the bulk properties of the system, yield a complete characterization of the topological phases. In a future perspective, going beyond the framework of static effective models, we plan to exploit the edge quantum mutual information and the edge squashed entanglement to investigate the fate of topological order in the full nonequilibrium dynamics of open quantum many-body systems.

\section*{Acknowledgements}
F.I. acknowledges support by MUR (Ministero dell’Università e della Ricerca) via the project PRIN 2017 ”Taming complexity via QUantum Strategies: a Hybrid Integrated Photonic approach” (QUSHIP) Id. 2017SRNBRK.

\appendix
\section{Hamiltonians, topology and symmetries}
\label{AppA}

The ten-fold classification of topological superconductors and insulators has been first discussed by Altalnd and Zirnbauer\cite{PhysRevB.55.1142} for spinful systems and subsequently applied also to spinless particles. It allows to identify the topological order and the number of edge modes according to the spatial dimensionality and the simultaneous presence/absence of particle-hole symmetry, time reversal symmetry and chiral symmetry.

The original Kitaev chain model \cite{Kitaev_2001} can be obtained by Eq. \ref{KCcurrent} with $q=0$. Due to the simultaneous presence of the three discrete symmetries listed above, it belongs to the BDI class of the Cartan classification with $\mathcal{Z}$ index.  The topological invariant is sensitive to the number $m$ of edge modes, with $m \in \{0$, $\pm1$, $\pm2$, $\dots\}$. However, for a one-dimensional (single-orbital) chain, it can only assume values $1$ or $0$, labelling respectively the topological and trivial phase. In general, a $\mathcal{Z}$ topological invariant in one dimension can be expressed by the winding number \cite{Maiellaro2018TopologicalPD}, even though the same phase diagram can be obtained by means of the Pfaffian invariant \cite{Maiellaro2018TopologicalPD}. Indeed, being the Hamiltonian in the Majorana basis ($H_M$) an antisymmetric matrix, the Pfaffian is a well defined quantity, $Pf[i H_M(k)]=-\mu-2t \cos(k)-2i \Delta \sin(k)$. Hence, the sign of the product of Pfaffians for $k=0$, $\pi$ switches at the gap closing points of the BdG band structure and thus the topological phase diagram can be computed by introducing the simple topological invariant $Q$ that reads
\begin{equation}
Q=Sign[(-\mu+2t)(-\mu-2t)] \, . 
\end{equation}
The presence of symmetries acting on the Kitaev chain is highlighted by resorting to the momentum representation. In this representation, the Hamiltonian reads:
\begin{eqnarray}
	\label{BdGKspaceKC}
	\tilde{H}(k)=\left(
	\begin{array}{cc}
		-2t \cos(k)-\mu&2i \Delta \sin(k)\\
		-2i \Delta \sin(k)&2t \cos(k)+\mu\\
	\end{array}
	\right).
\end{eqnarray}
As already mentioned, due to the superconducting order, the system fulfills the particle-hole symmetry that exchanges creation and annihilation operators, i.e. in second quantization language $c_j \leftrightarrow c_{j}^\dagger$. This symmetry operator, in momentum representation can be expressed by $P= \sigma_x K$, whose action on the Hamiltonian is
\begin{eqnarray}
	P H(k)P^{\dagger}=-H(-k) \, ,
\end{eqnarray}
where $K$ is the complex conjugation operator. Given a solution with energy $E$ and momentum $k$, the particle-hole symmetry ensures the presence of a solution with energy $-E$ and momentum $-k$.

Another symmetry condition satisfied by the system is invariance under time reversal. In the language of second quantization this means that time reversal leaves the creation and annihilation operators unaffected while it implements complex conjugation of all the complex-valued parameters: $(c_j,\ c_j^{\dagger}) \rightarrow (c_j,\  c_j^{\dagger})$, $i \rightarrow -i$. For spinless systems, time-reversal symmetry represents a symmetry condition for all the real-valued matrices. It is straightforward to show that in the chosen basis it coincides with the operator of complex conjugation: $T=K$, so that
\begin{eqnarray}
	T H(k) T^{\dagger}=H(-k) \, .
\end{eqnarray}
Finally, we can define the chiral symmetry as $\mathcal{C}=P T= \sigma_x$, whose action is
\begin{eqnarray}
	\mathcal{C} H(k) \mathcal{C}^{\dagger}=-H(k).
\end{eqnarray}
When currents are introduced, i.e. setting $q \neq 0$ in the generalized Kitaev model in Eq. \ref{KCcurrent}, the time-reversal symmetry is broken  ($(c_j, c_j^{\dagger}) \rightarrow (c_j,c_j^{\dagger})$, $t e^{iq} \rightarrow t e^{-iq}$), since the hopping strength is a complex-valued quantity. As a consequence, chiral symmetry is also broken, while particle-hole symmetry is preserved. As the current breaks two symmetries, it leads the system to the Cartan $D$ class of the ten-fold classification with the topological invariant corresponding to a $\mathcal{Z}_2$ index with only two distinct topological phases. The topological invariant $Q$ can now be expressed as:
\begin{eqnarray}
Q=Sign[Sign[Pf [i H_M(0)] Pf [i H_M(\pi)]]+Sign[q - q_c]] \, , \nonumber
\label{topoinariant}
\end{eqnarray}
where $Pf[i H_M(0/\pi))]\!\!=\!\!-\mu \pm 2t \cos(q)$ and $q_c= \arcsin(\Delta/t)$.
Similarly to the case of the unperturbed Kitaev chain, the topological invariant $Q$ provides a dichotomic topological label classifying the gap closing points. Indeed, when $q$ is smaller than the critical value $q_c$, gap closing points can only occur for $k=0$, $\pi$ and the sign of the Pfaffians product match topological/trivial phases of the system. On the other hand, for $q\geq q_c$, gap closing points no longer exist and are replaced by crossing points. These zero-energy band crossings correspond to trivial phases of the system.

\section{Spin representation of topological superconductors under a uniform particle current}
\label{AppB}

The Jordan-Wigner transformation \cite{Franchini2017} is a highly nonlocal mapping between fermionic and spin $1/2$ operators. On each site, an empty state is mapped into a spin up and an occupied one to a spin down. The nonlocal part of this mapping is called the Jordan-Wigner string and fixes the (anti)commutation relations between operators acting on distinct sites, by counting the parity of flipped sites to the left of the spin on which it acts. 

This transformation explicitly breaks the translational invariance of the model, by singling out a particular site as the initial point of the string. Denoting by $c_j$ and $c_j^\dagger$ the generic annihilation and creation fermionic operators, the Jordan-Wigner mapping is defined as follows:

\begin{eqnarray}
	c_{j} &=& e^{-i \pi \sum_{l=1}^{j-1} c^\dagger_{l} c_{l}}\sigma^{+}_{j} \, ,\\
	&& \nonumber \\
	c^\dagger_{j} &=& \sigma^{-}_{j}e^{i \pi \sum_{l=1}^{j-1} c^\dagger_{l} c_{l}} \, , \\
	&& \nonumber \\
	n_{j} &=& \frac{1-\sigma^{z}_{j}}{2} \, ,
	\label{JWtransformation}
\end{eqnarray}
where $j$ singles out the explicit lattice site. The aforementioned parity string of the overturned sites is $e^{-i \pi \sum_{l=1}^{j-1} c^\dagger_{l} c_{l}}$.
\begin{widetext}
The operators $\sigma_{j}^{\pm}=(\sigma_{j}^{x}\pm i\sigma_{j}^{y})/2$ are the well-known linear combinations of Pauli matrices and the last relation in Eq. (\ref{JWtransformation}) allows to express the parity operator of the fermionic site $j$ as $e^{-i \pi c^\dagger_{j} c_{j}}=\sigma_{j}^{z}$. Using the algebra of spin $1/2$ operators and observing that Pauli matrices acting on different sites commute, it is straightforward to derive the following spin-$1/2$ representation of the Kitaev chain in the presence of a particle current:
\begin{eqnarray}
	H_{spin} =  \frac{1}{2} \sum_{j=1}^{L-1} \biggl[\omega_q^-\sigma^x_{j} \sigma^x_{j+1}-\omega_q^+\sigma^y_{j} \sigma^y_{j+1}+ \sin(q) \biggl(\sigma^y_{j} \sigma^x_{j+1}-\sigma^x_{j} \sigma^y_{j+1}\biggl)\biggr] +  \frac{\mu}{2}\!\sum_{j=1}^{L} \sigma^z_{j},
	\label{KCcurrentSPIN}
\end{eqnarray}
\end{widetext}
where $\omega_q^{\pm}=\Delta\pm t \cos(q)$. We see that the fermionic model transforms into a $XY$ spin chain with a term mixing the $X$ and $Y$ components of the spins and a transverse external magnetic field along the $Z$-direction.
\bibliography{Bib}
\end{document}